\documentclass[aps,prd,superscriptaddress,showpacs,nofootinbib,twocolumn,floatfix]{revtex4-1}
\usepackage{mciteplus}
\usepackage{latexsym}
\usepackage{amsthm}
\usepackage{amsmath}
\usepackage{amssymb}
\usepackage{hepunits}
\usepackage{hyperref}
\usepackage{bbm}
\usepackage{bm}
\usepackage{xfrac}
\usepackage{color}
\usepackage{colordvi}
\usepackage{comment}
\usepackage{dcolumn}
\usepackage{times,latexsym,graphicx,wrapfig,url}
\usepackage{epsfig,lineno,bm}
\usepackage{subfigure}
\usepackage{slashed}

\newcommand{\be}{\begin{eqnarray}}
\newcommand{\ee}{\end{eqnarray}}
\newcommand{\bea}{\begin{eqnarray}}
\newcommand{\eea}{\end{eqnarray}}

\newcommand{\bef}{\begin{figure}[htbp]\begin{center}}
\newcommand{\eef}{\end{center}\end{figure}}

\newcommand\FNAL{Fermi National Accelerator Laboratory, Batavia, IL, 60510, United States}

\def\lsim{\mathrel{\rlap{\lower4pt\hbox{\hskip1pt$\sim$}}
    \raise1pt\hbox{$<$}}}
\def\gsim{\mathrel{\rlap{\lower4pt\hbox{\hskip1pt$\sim$}}
    \raise1pt\hbox{$>$}}} 

\begin{document}
{\scriptsize FERMILAB-PUB-15-550-A} \\  \\ \\

\title{ Probing Light Thermal Dark-Matter With a Higgs Portal Mediator}
\author{Gordan~Krnjaic}          \thanks{krnjaicg@fnal.gov}  \affiliation{\FNAL}
\date{\today}


\begin{abstract}

We systematically study light ($<$ few GeV) Dark Matter (DM) models that thermalize with visible
matter through the Higgs portal and identify the remaining gaps in the viable parameter space. 
Such models require a comparably light scalar mediator that mixes with the Higgs to avoid DM overproduction and 
can be classified according to whether this mediator decays (in)visibly. In a representative benchmark  model with 
Dirac fermion DM, we find that, even with conservative assumptions about the DM-mediator coupling and mass ratio, the regime in which the mediator is  heavier than the DM is fully ruled out by  a combination of collider, rare meson decay, and direct detection limits;  future and planned experiments including NA62 can further improve sensitivity to scenarios in which the Higgs portal interaction does not determine the DM abundance.  
 The  opposite  regime in which the mediator is lighter than the DM and the latter annihilates to pairs of visibly-decaying mediators is still viable, but much of the parameter space is 
 covered by rare meson decay, supernova cooling, beam dump, and direct detection constraints. Nearly all of these conclusions apply broadly to the simplest variations (e.g. scalar or asymmetric DM). Future experiments including  SHiP, NEWS, and Super-CDMS SNOLAB can 
greatly improve coverage to this class of models.

\end{abstract}

\maketitle


\section{Introduction}
Although evidence for the existence of Dark Matter (DM) is overwhelming, its particle identity remains unknown and discovering its short distance properties is a top priority in fundamental physics. 
This task is especially daunting because 
viable DM candidate masses  span dozens of orders of magnitude with different cosmological histories
 and phenomenological consequences.
However, if dark and visible matter achieve thermal equilibrium in the early universe,  the viable mass range is much narrower, $m_{\rm DM} \sim \keV-100  \> \TeV$; 
below a few $\keV$, DM is too hot for structure formation and  above $\sim 100$ TeV, DM is in tension with perturbative unitarity.  Thermal contact also generically overproduces DM in the early universe, so such scenarios require a depletion mechanism to yield the observed abundance. This feature motivates appreciable non-gravitational interactions with visible matter and serves as a well motivated and largely model-independent organizing principle for the broad DM discovery effort. 

For the upper half of the thermal window, $m_{\rm DM} \sim$ GeV--100 TeV, DM can carry 
electroweak quantum numbers and annihilate via Standard Model (SM) interactions. 
For the lower half,  $m_{\rm DM} \sim$ keV--GeV,  thermal dark matter with weak-scale (or weaker) interactions  
is overproduced in the early universe \cite{Lee:1977ua}, so viable scenarios 
require additional, SM neutral mediators to deplete the overabundance. Unless SM fields are charged directly under additional
forces (e.g. $U(1)_{B - L}$), these mediators will mix with the SM through at least one of the  renormalizable ``portal" operators  
\be 
H L  ~~,~~ B^{\mu\nu}~~,~~  H^\dagger H ~,~
\ee
where $L$ is a lepton doublet, $B^{\mu\nu}$ is the hypercharge field strength tensor, and $H$ is
the Higgs doublet. 

Stable thermal DM interacting through the lepton portal, $HL$, is difficult to 
engineer because either the DM decays through this interaction (e.g. DM is a right handed or sterile neutrino) or the mediator is fermionic and 
SM-DM scattering is proportional to neutrino masses so the two sectors never thermalize (see \cite{Shakya:2015xnx} for a review). 
Vector mediators can kinetically mix with $B^{\mu\nu}$ and there is a growing 
 effort to test this scenario \cite{Izaguirre:2015yja,Essig:2013lka,Essig:2010xa,Merkel:2011ze,Abrahamyan:2011gv,Izaguirre:2013uxa,Dharmapalan:2012xp,Izaguirre:2015pva,
Izaguirre:2014bca,Batell:2014mga,Kahn:2014sra,Battaglieri:2014qoa,Hochberg:2015pha,Curtin:2014cca}.

In this paper we  study a simple class of light ($<$ few-GeV) thermal DM models with a singlet scalar mediator that mixes with the SM through the Higgs portal. 
For a singlet mediator $\Phi$, the mixing arises from the renormalizable operators 
\be\label{eq:mixing-lagrangian}
{\cal L}_{\Phi,H} =  (   A_{\tiny \Phi H} \Phi  +  \lambda_{\Phi H} \Phi^2 )      H^\dagger H~.
\ee
\noindent After electroweak symmetry breaking, diagonalizing the scalar mass terms that arise from Eq.~(\ref{eq:mixing-lagrangian}) yields mass
eigenstates $\phi$ and $h$, which we identify as the DM-SM mediator and the Higgs boson, respectively (see Appendix A for a discussion).
Our representative benchmark scenario consists of a  Dirac fermion DM candidate coupled to the mediator $\phi$ via 
\be\label{eq:phiDMlag} 
{\cal L}_{\phi,{\rm DM}} = \phi (g_{\chi}  \bar \chi \chi + g_{\chi}^\prime \bar\chi \gamma^5\chi) ~,~
\ee 
where $g_{\chi}$ and $g^\prime_{\chi}$ are the parity even and odd couplings, respectively. However, as we will see below, the relevant physics for light DM  is captured by the scalar interaction, so we will omit parity odd coupling $g_\chi^\prime$ from our benchmark without loss of essential generality (see discussion in Sec. \ref{sec:variations}).

Since $\Phi$ mixes with the scalar component of $H$ (the Higgs boson, $h$), it acquires a coupling to SM fermions,
which we parametrize with the mixing angle $\sin\theta$ and expand in the mass basis to obtain the mediator interaction
\be\label{eq:phiSMlag}
{\cal L}_{\phi,{\rm SM}} =   \phi  \sin\theta  \sum_f   \frac{m_f}{v}   \bar f f ~~,~~
g_{f}  \equiv  \frac{m_{f}}{v} \sin\theta ~,~~
\ee 
where $f$ is a SM fermion of mass $m_{f}$  and $v \simeq  246 \, \GeV$ is the SM Higgs vacuum expectation
value. Although this is only one of many scenarios for DM interacting through the Higgs portal, it captures much of the essential physics, so 
most of the constraints and projections will apply to a much broader class of variations on this simple setup.  


\begin{figure}[t] 
\includegraphics[width=7cm]{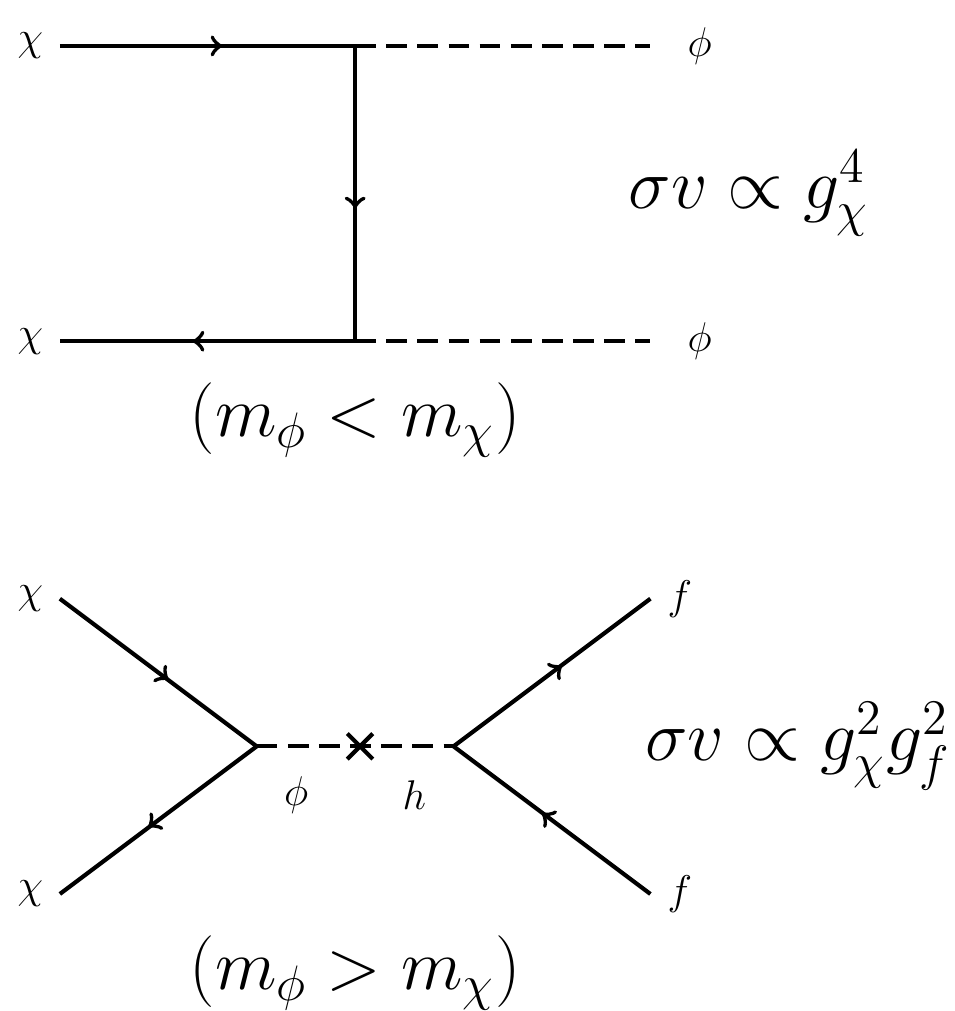} \hspace{5cm}
  \caption{Leading Feynman diagrams giving rise to $\chi$ annihilation in the early universe. If $m_\chi > m_\phi$ the annihilation is predominantly through the $t$-channel and the mediator decays  into SM states via Higgs mixing. If $m_\chi < m_\phi$, DM annihilates directly to SM fermions through the $s$ channel which depends on the SM-mediator coupling and is the most predictive scenario; If $m_\phi >  2 m_\chi$  the $\phi$ will decay invisibly to dark matter.  In the $ 2 m_\chi > m_\phi > m_\chi$ regime, it may also be possible to annihilate through the forbidden channel \cite{D'Agnolo:2015koa}}
   \label{fig:AnnihilationDiagram}
\vspace{0cm}
\end{figure}

Light DM interacting through the Higgs portal has been considered before in the context of minimal DM coupled directly to
 the portal \cite{Burgess:2000yq}, as a byproduct of Higgs decays \cite{Pospelov:2011yp}, as pair produced in rare $B$ and $K$ decays \cite{Bird:2006jd},
as coupled to a scalar mediator mixed with the Higgs \cite{Pospelov:2007mp,Bird:2004ts,Schmidt-Hoberg:2013hba}, as a sub-eV 
non thermal candidate \cite{Piazza:2010ye}.  The bounds on a light, Higgs portal scalar in the context of self-interacting DM were considered
in  \cite{Kouvaris:2014uoa,Kainulainen:2015sva}. This paper adds to the literature by carefully computing the relic density of DM through a highs-mixed mediator
 including the effects of hadronic final states; updating constraints in light of recent direct detection, LHC, and rare meson decay results; and 
 discussing the implications for the simplest DM variations (e.g asymmetric, inelastic, scalar).  
We find that for heavier mediators $m_\phi > m_\chi$, DM annihilating directly 
into SM particles is  ruled out for nearly all DM candidates under the most conservative assumptions regarding the DM-mediator couplings and mass ratios. We also find that when the mediator is lighter and the relic abundance is set by secluded annihilation $\chi \bar \chi \to \phi \phi$, the mediator-Higgs mixing is bounded from below by the DM thermalization requirement and there remains much viable parameter space. Finally, we identify a representative set of future direct detection and meson decay experiments to extend coverage 
to much of the remaining territory. 

This paper is organized as follows: In Section \ref{sec:thermalrelic} we compute the DM relic density and discuss how to conservatively compare this target against different kinds of bounds; in Section \ref{sec:gen}, we describe generic constraints and future experimental projections applicable to the entire parameter space; in Sections \ref{sec:invisible} and  \ref{sec:visible}  we specify 
to the regimes in which the mediator decays to the DM and SM respectively; In Section \ref{sec:compressed} we discuss the unique features of the compressed
region of parameter space in which $m_\chi < m_\phi < 2m_\chi$;  in Section  \ref{sec:variations} we outline how varying the assumptions about the DM candidate relative to our benchmark scenario (introduced above) changes the viable parameter space; finally in 
Section \ref{sec:conclusion} we offer concluding remarks.

\medskip 


\begin{figure*}[t] 
\vspace{0.1cm}
\hspace{-0.7cm}
\includegraphics[width=8.7cm]{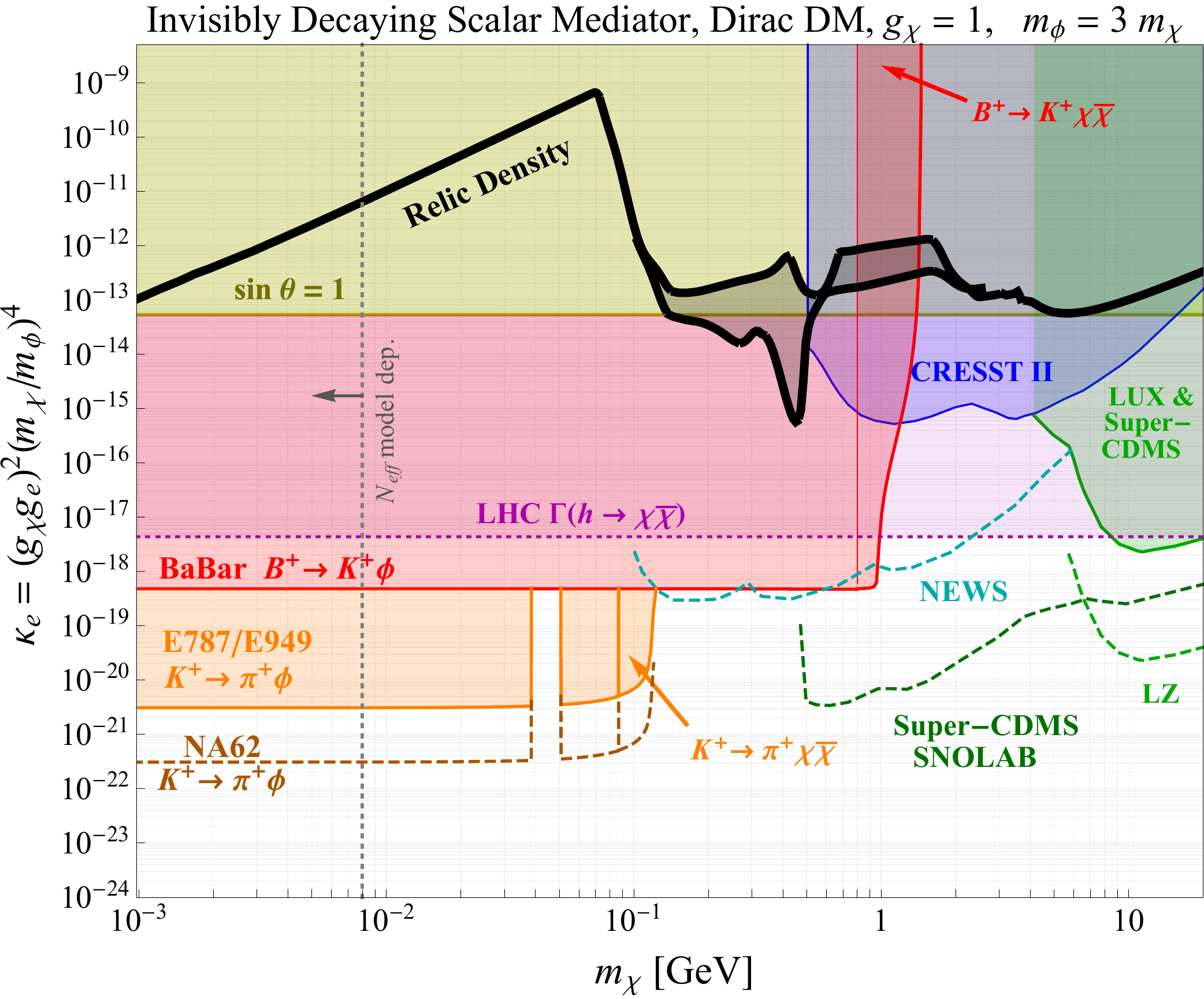}  ~~
\includegraphics[width=8.7cm]{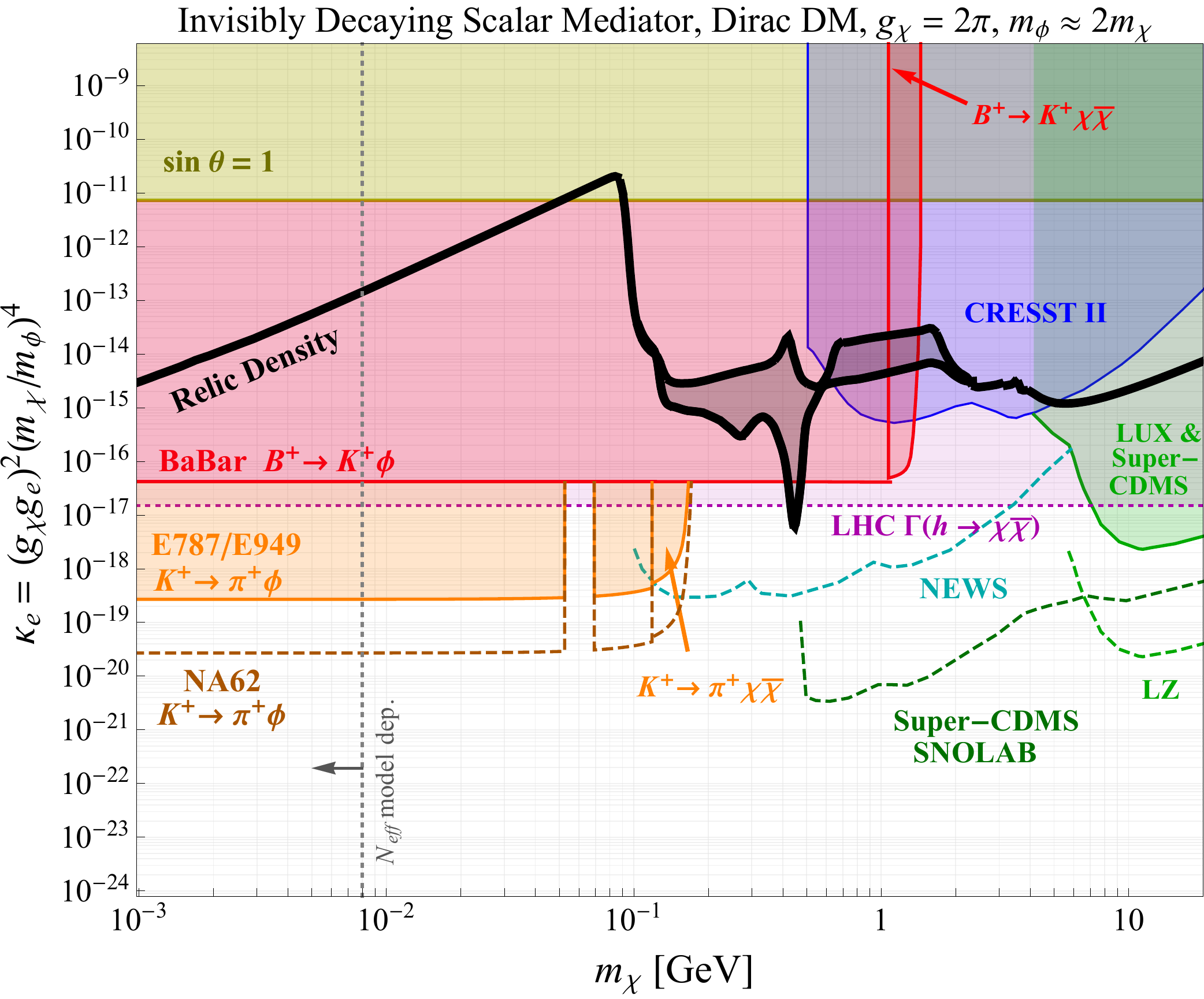}  ~~
\hspace{-0.7cm}
\includegraphics[width=8.7cm]{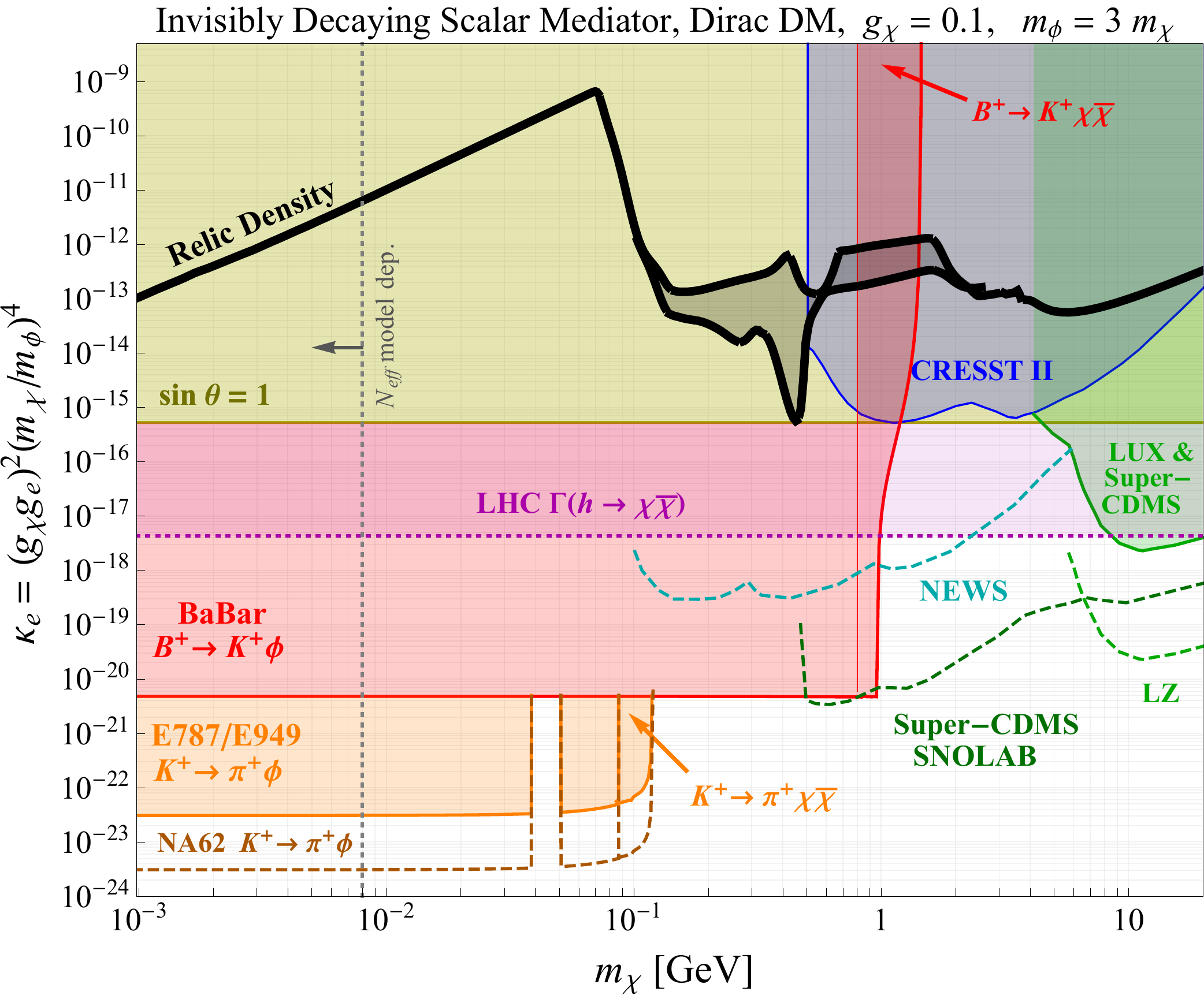}  ~~
\includegraphics[width=8.7cm]{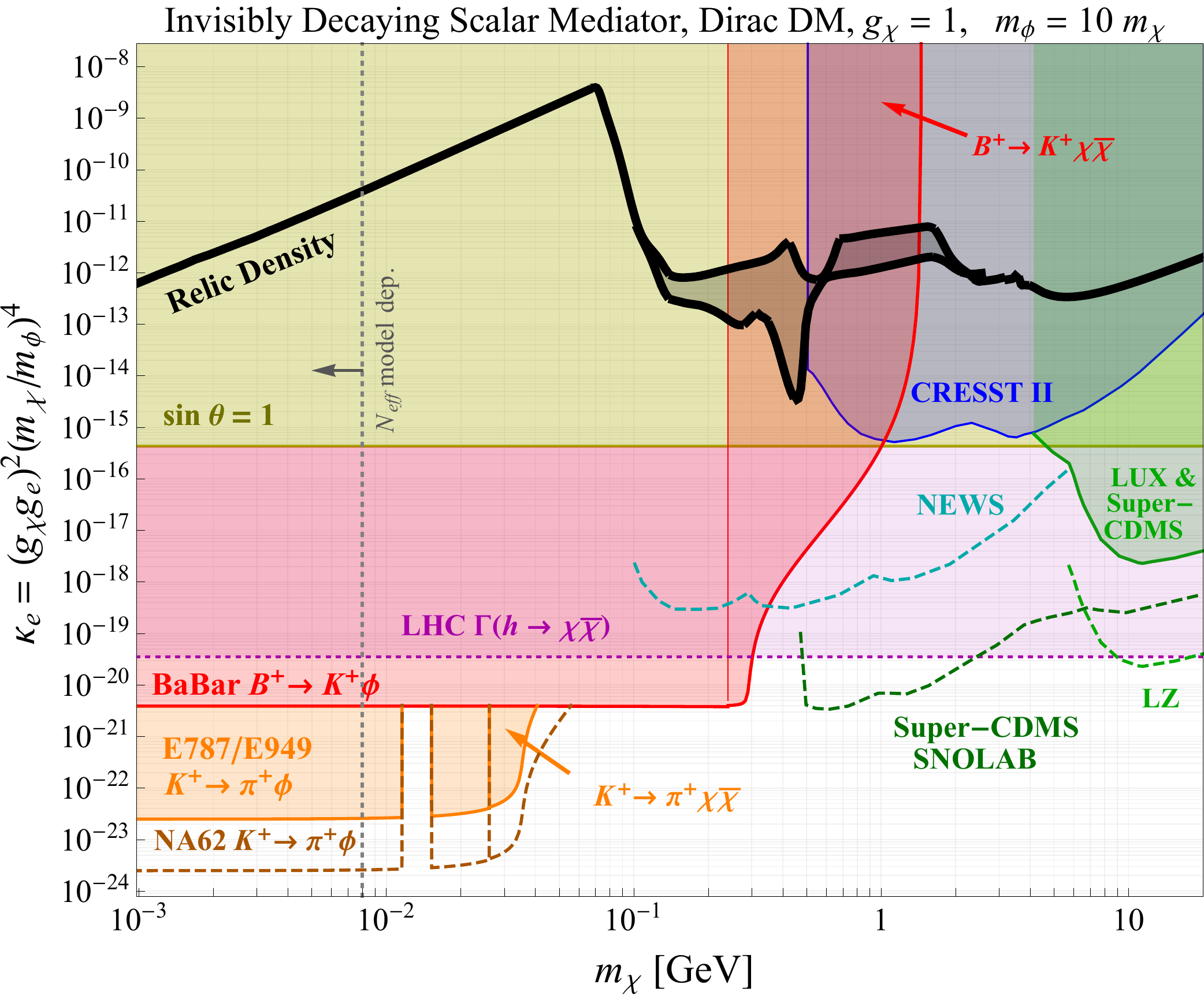}  ~~

  \caption{Experimental constraints on Dirac fermion DM that annihilates through a light, Higgs-mixed mediator. We normalize the vertical axis using the $e$-$\phi$ coupling, $g_e$ introduced in the text because this coupling always contributes to the annihiation over the mass range considered here-- see discussion in Section II.
 {\bf Top Left: } Parameter space for $m_\chi < m_\phi$ compared against the relic density contour computed assuming $m_\phi = 3 m_\chi$ (solid black curve). The 
 curve bifurcates near $m_\chi \sim m_\pi$ where there is disagreement in the literature about light Higgs couplings to hadronic states (see text).   
Like the relic density contour, the direct detection constraints are also invariant under different assumptions about the mass ratio and DM-mediator coupling since
the SM-DM scattering cross section is proportional to the $\kappa_e$ variable plotted on the vertical axis. 
However, for meson decay and collider constraints, which only constrain
the mediator-Higgs mixing, we adopt the conservative values $g_\chi = 1$ and $m_\chi/m_\phi = 1/3$ for building $(g_\chi g_e)^2 (m_\chi/m_\phi)^4$ for comparison with the 
solid black relic curve; choosing
smaller values of either quantity makes these constraints stronger -- except in the resonant annihilation region. {\bf Top Right: } Same as left, but in the resonant annihilation region $m_\phi \approx 2 m_\chi$, which
is the only regime in which the relic density curve moves appreciably. This plot also adopts the extreme value $g_\chi = 2\pi$ near the perturbativity limit, and reveals the maximum amount of viable parameter space for this scenario. As on the top-left plot, direct detection constraints and projections remain invariant, but the meson and collider bounds shift slightly as they are now computed for $m_\chi/m_\phi  = 1/2.2$  instead.
{\bf Bottom Right: } Same as top-left, but with $m_\phi = 10 m_\chi$. {\bf Bottom Left: } Same as top-left, but with the reduced coupling $g_\chi = 0.1$.
  }
   \label{mainfig}
\vspace{0cm}
\end{figure*}



\section{\bf Thermal Relic Comparison} \label{sec:thermalrelic}



\noindent { \bf  ~~~~~~~~~~~~~~~~~~~~~~~~     Direct Annihilation ($ m_\chi < m_\phi$) }  \\

In the regime where the mediator is heavier than the DM, the annihilation can only proceed via direct annihilation to SM fermions through the $s$-channel.\footnote{For an interesting counterexample see \cite{D'Agnolo:2015koa} where
DM annihilates predominantly to pairs of heavier mediators (the so-called ``forbidden" channel) by sampling the tail of the DM Boltzmann distribution at freeze out. For completeness, we also mention the possibility of $2\to3$ \cite{Rajaraman:2015xka} and 3 $\to$ 2 annihilation \cite{Hochberg:2014dra} though the cases studied in these papers represent departures from the Higgs-mixing benchmarks considered in this paper}
To leading order, the annihilation rate for Dirac fermion annihilation into elementary fermions $\chi \bar \chi \to \bar f f$ is $p$-wave  
\be \label{eq:directannihilation}
\sigma v_{\rm rel.}(\chi\chi \to f \bar f) &=&  \frac{  g_\chi^2 g_f^2  m_\chi^2 v_{\rm rel.}^2     }{8\pi(m_\phi^2 - 4 m_\chi^2)^2}   \propto g_\chi^2 g_f^2  \left( \frac{m_\chi}{m_\phi} \right)^{\!4} \frac{1}{m_\chi^2}~~
, ~~~~~~~
\ee
where $v_{\rm rel.}$ is the relative velocity between annihilating particles. Away from resonance at $m_\phi \sim 2m_\chi$ (and up to corrections of order $m^2_\chi/m^2_\phi)$, for a fixed value of $g_\chi^2 g_f^2  \left( m_\chi /m_\phi\right)^{\!4}$,  the annihilation rate is independent of the $m_\chi/m_\phi$ ratio or the individual values of $g_\chi$ and $g_f$.
From the parametric dependence in Eq.~(\ref{eq:directannihilation}), it is convenient to define a dimensionless quantity 
\be
\kappa_f \equiv  g_\chi^2 g_f^2   \left( \frac{m_\chi}{m_\phi} \right)^{\!4} =   g_\chi^2 \left(     \frac{ m_f }{v}  \sin \theta \right)^2 \left( \frac{m_\chi}{m_\phi} \right)^{\!4}~,~~
\ee
so that the annihilation rate $\chi \chi \to \phi^* \to f f$  is uniquely specified by the value of $\kappa_f$ for a given  value of $m_\chi$.
In the regime where annihilation is predominantly to electrons, achieving the observed relic abundance requires  
\be \label{eq:kappa-freezeout}
\kappa_e \simeq   10^{-11}    \left(  \frac{ 0.1}{\Omega_\chi h^2 } \right)   \left( \frac{m_\chi}{ 10\, \MeV} \right)^2~.~
\ee
 Including all kinematically accessible channels and exploiting the mass 
proportionality of Higgs couplings, the full annihilation cross section can be written in terms of  $\kappa_e$   
\be
\sigma v_{\rm rel.}(\chi \chi \to {\rm SM}) \propto     \frac{1}{m_\chi^2} \sum_f \kappa_f  =  \frac{\kappa_e}{m_\chi^2}  \sum_f  \left( \frac{m_f}{m_e} \right)^2 ~,~~~
\ee
which  is applicable to all $m_\chi$ (MeV--GeV)  considered in this paper, so we will present our direct annihilation results in terms of $\kappa_e$ without
loss of generality. For a more careful treatment of thermal freeze out, corresponding to the methodology in our numerical studies, see Appendix B. 

For $m_\chi \gsim \Lambda_{\rm QCD}$, the 
annihilation also proceeds through several hadronic channels, whose interactions  with the mediator are not simply-related to quark Yukawa couplings
(e.g. $\chi \chi \to \pi^+ \pi^-$). To account for these final states, we extract this coupling 
from simulations of hadronically-decaying light-Higgs bosons \cite{Clarke:2013aya} with the ansatz 
\be \label{eq:hadronic-phi-coupling}
  g_{f}(s) \simeq   \sin\theta \sqrt{   \frac{8 \pi}{m_h} \Gamma(h \to {\rm SM})  }  ~ \biggr  |_{m_h = \sqrt{s}}~~~,
\ee
which is valid up to corrections of order $m^2_f / s$ where $s$ is the mandelstam variable ($s\approx 4m_\chi^2$ near freeze out). However, there are significant, order of magnitude discrepancies among the various
computations of  $\Gamma_h(h \to {\rm hadrons})$ \cite{Gunion:1989we,Voloshin:1985tc,Grinstein:1988yu,Raby:1988qf,Donoghue:1990xh,Truong:1989my}  (see Fig. 1  in \cite{Clarke:2013aya} ), so we regard
this ansatz as reasonable in the interpolation region. We have checked that the couplings extracted using this approach recover
the correct light Higgs width in the regions well above and below $\Lambda_{\rm QCD}$. 

The solid black lines shown in the panels of Fig. 2 represent the relic 
density contour computed precisely for different choices of mass and coupling ratios. These plots illustrate
how, away from resonance, the parameter space for thermal freeze-out is invariant with respect to $\kappa_e$ as a function of $m_\chi$.
The curves bifurcate near $\Lambda_{\rm QCD}$ to account for the different theoretical extractions of hadronic couplings in the region where 
annihilation is predominantly to hadrons; the upper and lower curves bracket the range of discrepant values from the literature \cite{Gunion:1989we,Voloshin:1985tc,Grinstein:1988yu,Raby:1988qf,Donoghue:1990xh,Truong:1989my}.

In analogy with the $s$-channel annihilation process, the  cross-section for non-relativistic DM-SM scattering in direct detection through $\phi$ exchange 
also has the same scaling $\sigma \propto \kappa_e/m_\chi^2$, up to order one corrections
when the DM-nucleon masses are comparable.  However, the constraints at most accelerator based experiments do not scale in this way; typically they only constrain 
the magnitude of a SM-mediator coupling and are insensitive to the mediator-DM coupling (and often also the masses of DM and mediator when the energy scale of the process is larger than the masses of dark sector states). 
For instance, the $B$ and $K$ decay bounds (See Sec IV) depend on the mass ratio since these bounds only constrain the mediator-SM coupling, so we have to choose a value of $g_\chi^2(m_\chi/m_\phi)^4$ 
to construct $\kappa_e$ and plot these constraints against the thermal relic contour. For these bounds, choosing a very small $g_\chi$ or $m_\chi/m_\phi$ ratio trivially covers 
arbitrarily small values of $\kappa_e$ so the conservative choice is to take $g_\chi \sim m_\chi/m_\phi \sim {\cal O}(1)$
(see the caption in Fig. 2 and discussion below for more details).

 On the upper-right panel of Fig. 2, we evaluate the same relic target, experimental bounds, and future projections
 for   $m_\phi/m_\chi = 2.2$, which is very close to resonance, for which the relic target does move appreciably downwards relative to the relic density curves in other panels. In this panel we also adopt the extreme value $g_\chi = 2 \pi$ near the perurbativity limit, for which this model requires a UV completion
 near the GeV scale. Nonetheless,
 even in this {\it extremely} conservative regime where the constraints are the weakest (all either shift upwards or are unaffected relative to the other panels) and thermal freeze requires the smallest $\kappa_e$ values for the direct annihilation scenario, 
 there is no viable territory left and the direct annihilation scenario is ruled out.\footnote{This conclusion depends somewhat on how large
 of a coupling a  given model can viably generate and on the theory uncertainty of the mediator-SM coupling near $\Lambda_{\rm QCD}$, but
 aside from these caveats, the direct annihilation scenario is ruled out. } From the other plots in Fig. 2, we see that the relic target for direct annihilation 
 is decisively  covered over the full MeV--GeV mass range.


\bigskip 
\noindent { \bf  ~~~~~~~~~~~~~~  Annihilation Into Mediators ($ m_\chi > m_\phi$) }  \\

If the mediator is lighter than the DM,  the  direct annihilation annihilation is sharply suppressed relative to the  t-channel process $\chi \chi \to \phi \phi$, 
which no longer scales as $\kappa_e /m_\chi^2$. Instead we have 
\be \label{eq:tchannelannihilation}
\sigma v_{\rm rel.}(\chi \chi \to \phi \phi)  = \frac{3 g_\chi^4 v_{\rm rel}^2}{ 128 \pi m_\chi^2}~,~
\ee
which is independent of the $\phi$-$h$ mixing angle, so thermal freeze out is compatible with a wide range
of mixing angle values.  However, for sufficiently small  $\sin \theta$, the dark and visible sectors no longer thermalize, so there is a lower 
bound on the parameter space; for smaller mixing angles the abundance must be generated by a nonthermal mechanism, so this scenario is beyond 
the scope of this work.


\section{Generic Bounds} \label{sec:gen}

\label{sec:generic}

\medskip
\noindent\noindent{\bf CMB}  
Although any viable thermal DM candidate is frozen out well before recombination, out of equilibrium annihilation 
 around $z\sim 1100$ can still reionize hydrogen at the surface of last scattering and
 thereby modify the CMB power spectrum.  The Planck constraint $\langle \sigma v \rangle_{\rm cmb}/m_\chi \lsim 3 \times 10^{-28}\, \cm^3 {\rm s}^{-1}  \GeV^{-1}$ rules out thermal DM below 10 GeV  \cite{Ade:2015xua}  if the annihilation rate is $s$-wave (velocity independent). However, the annihilation rates for our benchmark scenarios in Eqs.~(\ref{eq:directannihilation}) and ~(\ref{eq:tchannelannihilation}) are $p$-wave so  the annihilation rate is many orders of magnitude smaller on account of the velocity redshift from $T\sim m_\chi$ at freeze out
and $T\sim \eV$ at recombination. Thus neither of our annihilation topologies is constrained by CMB power injection.


 \medskip
 \noindent\noindent{\bf Relativistic Degrees of Freedom } 
 If DM freezes out after neutrinos decouple, the annihilation byproducts can reheat photons relative to neutrinos, 
 thereby decreasing the effective relativistic degrees of freedom, $N_{\rm eff.}$ \cite{Nollett:2013pwa}. At face value this 
 excludes thermal DM below $\lsim 10$ MeV (with order one variations depending on the particle identity of the DM). However, this bound is 
 model dependent because the deficit in $N_{\rm eff.}$ can be compensated for with additional hidden sector radiation or new physics in the neutrino sector.


\begin{figure}[t] 
\vspace{0.1cm}
\hspace{-0.1cm}
\includegraphics[width=7.5cm]{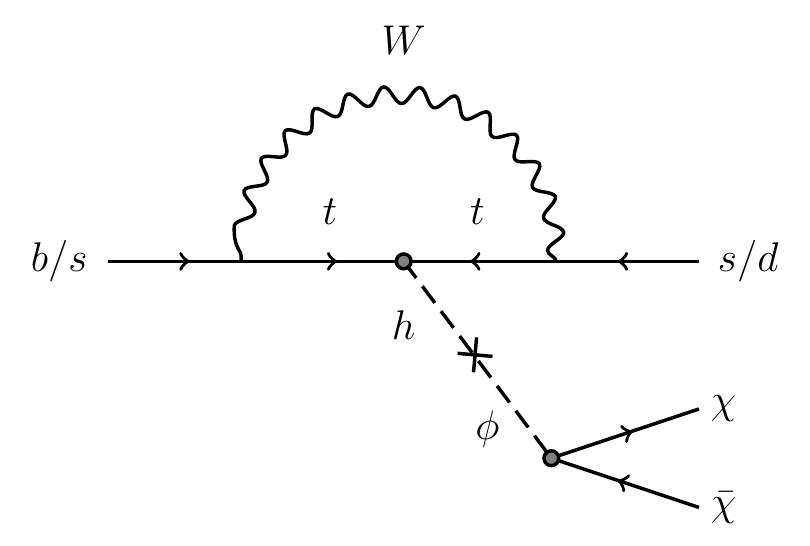}  ~~
  \caption{Leading short distance contribution to $B^+ \to K^+ \chi \chi$ and $K^+ \to \pi^+ \bar \chi \chi$ decay due to scalar mediated interactions. For $m_\phi < m_{B} - m_{K}$, this decay can also proceed via $B^+ \to K^+ \phi$  Similar diagrams yield for $\phi$ mediated contributions to fully SM final states  (e.g. $B^+ \to K^+ \mu^+ \mu^-$). }
   \label{fig:mone}
\vspace{0cm}
\end{figure}



\medskip
\noindent\noindent{\bf Higgs Decays } 
The $\phi$-$h$ mixing induces an invisible decay channel for the SM Higgs via $h \to \chi \chi$. 
Assuming only SM Higgs production mechanisms, the strongest limit on the invisible width arises 
from LHC measurements of $pp \to {\rm ~ jets} + \displaystyle{\not}{E}_T$, which can be interpreted 
to constrain the VBF production mechanism $pp \to {\rm ~ jets} + (h \to \chi \chi)$. A recent
ATLAS measurement has extracted a limit of ${\rm Br}(h \to {\rm invisible}) < 0.3$  \cite{Aad:2015txa}. which
for our scenario implies
\be
 g_\chi^2 \sin^2\!\theta \lsim  4 \times 10^{-5}~~,~~
\ee
or in terms of the variable plotted in top left panel of Fig. 2, $\kappa_e \lsim  7 \times 10^{-18}$,
 where the mass ratio is conservatively taken to be $m_\chi /m_\phi = 1/3$; heavier mediators make this constraint
 more severe, so this choice reveals the available gaps subject to the condition that the mediator decays invisibly and that $\chi \chi \to ff$ annihilation is off resonance.
 
 In addition to the mixing, the mixed $\phi-h$ quartic interaction may also contribute to exotic Higgs decays via $h \to \phi \phi$  \cite{Curtin:2013fra}. 
  If $\phi$ decays invisibly to DM, this process contributes to the Higgs invisible width, and if $\phi$ decays visibly the process can induce 
 an array of SM final states, which reconstruct the Higgs invariant mass and yield nested internal resonances. However, the bounds
 and prospects for both scenarios depend exclusively on the size of the quartic which does not affect the DM thermal history or the 
 bounds presented in this paper, so a proper treatment of this possibility is beyond the scope of the present work. 

We also note that there are additional constraints on the mixing angle $\sin\theta$  from rare $h\to \phi \phi$ decays. However, the branching ratio for this process depends on a different diagrams which are sensitive to the mixing angle, mixed $h^2 \phi^2$ quartic coupling, and the $\phi^3$ cubic coupling, so the precise bound arising from this process is model dependent and cannot be presented in Fig. 2 without additional assumptions about these other parameters.

\section{Invisibly Decaying Mediator $(m_\phi > 2m_\chi)$}
\label{sec:invisible}

\noindent{\bf Rare Meson Decays }  If $\phi$ decays invisibly, this scenario induces rare meson decays $B^+\to K^+ \phi$ and is constrained by limits on 
the $B^+ \to K^+ \nu \bar \nu$ branching fraction.  The loop level process arises from the effective 
Higgs mixing interaction \cite{Bird:2004ts,Bird:2006jd}
\be\label{eq:lagFCNC}
{\cal L}_{\rm FCNC} \supset ( C_{sb}  \bar s_L b_R   +  C_{sd}  \bar s_L d_R    )\phi~,~~~
\ee 
where $C_{sb, sd}$ are effective coefficients that induce flavor changing processes.


\bigskip
\noindent{\bf  B-Meson Decays }  For B-mesons, The effective coefficient of interest is  
\be
\label{eq:Csb}
C_{sb} &=&  \frac{3 g_W^2 m_{b} m_t^2 V_{ts}^* V_{tb}   \sin\theta  }{ 64 \pi^2 m_W^2 v }  = 6.4 \times 10^{-6} \sin\theta ~,~~~~
\ee
and this interaction has the partial width \cite{Dolan:2014ska}
\be \label{eq:fullbwidth} \hspace{-0.1cm} &&
\Gamma_{\! B\to K\phi} \!= \!\frac{|C_{sb}|^2  f_0(m_\phi)^2 \!}{16 \pi m_{B^+}^3}\! \left( \! \frac{m_{B^+}^2 - m_K^2}{m_b - m_s} \!\! \right)^2  \hspace{-0.2cm} \xi(m_{B}, m_{K}, m_\phi)  ,~~~~~~~
\\ \nonumber \\
&&   ~~~~~~~~~~~~~~~\xi(a,b,c) = \sqrt{   (a^2 - b^2 -c^2)^2 - 4 b^2 c^2 }~,
\ee
where the scalar form factor can be parametrized
$f_0(q)  = 0.33(1 - q^2/38 \, \GeV^2)^{-1}$ \cite{Ball:2004ye}.  The total $B$-meson width is $\Gamma_{B^+} = 4.1 \times 10^{-13}$ GeV \cite{Agashe:2014kda}, so the branching ratio has the approximate scaling
\be
{\cal B}r(B^+\to \! K^+\phi) &\sim& \frac{|C_{sb}|^2  f_0(m_\phi)^2 \!}{16 \pi} \frac{ m^3_{B^+} }{m_b^2 \Gamma_{B^+}} 
 \approx 1.5 \, \sin^2\!\theta  ,~~~~~~~~
\ee
which, for  our conservative benchmark inputs $g_\chi =1$ and $m_\chi = 3 m_\phi$, the BaBar limit ${\cal B}r(B^+\to \! K^+ \nu \bar \nu) < 1.6 \times 10^{-5}$ \cite{Lees:2013kla} requires
\be
\kappa_e = (g_\chi g_e)^2\left(\frac{ m_\chi}{ m_\phi}\right)^{\!4} \lsim 5.6 \times 10^{-19}~.
\ee
 The exact bound for this DM/mediator mass ratio shown in Fig. 2 (left) is computed from Eq.~(\ref{eq:fullbwidth}) using 
the efficiencies used in \cite{Lees:2013kla} is slightly stronger because the two-body $B^+\to \! K^+\phi$ process has greater kinematic acceptance relative to
 $B^+\to K^+\nu\bar \nu$.


\bigskip
\noindent{\bf   Kaon Decays }
An invisibly decaying light scalar can also yield $K\to \pi \phi$ decays for which the partial width is 
\be\label{eq:fullkwidth}
\hspace{-0.2cm}
\Gamma_{K^+\to \pi^+\phi} =  \!\frac{|C_{ds}|^2  }{16 \pi m_K^3}\! \left( \! \frac{m_{K^+}^2 \! - \!  m_{\pi^+}^2}{m_s \! - \! m_d} \!\! \right)^2 \hspace{-0.2cm} \xi(m_{K}, m_{\pi}, m_\phi),~~~~~~~
\ee
Unlike in Eq.~(\ref{eq:fullbwidth}), the analogous scalar form factor is close to unity \cite{Marciano:1996wy} and can  be neglected. The effective FCNC coefficient from Eq.~(\ref{eq:lagFCNC}) is 
\be
C_{sd} &=&  \frac{3 g_W^2 m_s m_t^2 V_{ts}^* V_{td}   \sin\theta  }{ 64 \pi^2 m_W^2 v }   = 1.2 \times 10^{-9} \sin\theta~,~~~
\ee
The total Kaon width is $\Gamma_{K^+} = 5.3 \times 10^{-17} \,\GeV$, so the branching ratio is approximately
\be
{\cal B}r(K^+\to \pi^+ \phi) \sim \frac{|C_{sd}|^2 }{16 \pi}  \frac{m^3_{K^+}}{m_s^2 \Gamma_{K+}} \approx  6.7 \times 10^{-3}\,  \sin^2\!\theta ~,~~~~~~~
\ee
This final state contributes to the E797 and E949  measurements of ${\cal B}r(K^+ \to \pi^+ \nu \bar \nu) =  (1.73^{+1.15}_{-1.05}) \times 10^{-10}$ \cite{Artamonov:2008qb}). 
To avoid an order one correction to  we demand $\sin^2\!\theta \lsim  2.5 \times  10^{-8}$, which, for our  benchmarks $g_\chi = 1$ and $m_\phi = 3 m_\chi$ implies
\be
\kappa_e = (g_e g_\chi)^2\left(\frac{ m_\chi}{ m_\phi}\right)^{\!4}  \lsim  1.4 \times 10^{-21}~.~
 \ee
 The exact bound computed using Eq.~(\ref{eq:fullkwidth}) and the efficiencies in \cite{Artamonov:2008qb} is presented as the orange regions of Fig. 2; the sensitivity gap near $m_\chi$ corresponds to an experimental cut on final-state pion momentum. Also shown in  Fig. 2 is the projected sensitivity of NA62 \cite{Ruggiero:2013oxa} assuming $10^{13} K^+$ and sensitivity for a 10 \% measurement of the $K^+ \to \pi^+ \nu \bar \nu$ branching
ratio with identical cuts and efficiencies as \cite{Artamonov:2008qb}. 


\medskip
\noindent\noindent{\bf Direct Detection } 
For light, non-relativistic DM with $v\sim 10^{-3}$, scattering off nuclei
yields recoil energies that fall below experimental thresholds. 
However, electron scattering results from XENON10 and
future results can constrain this parameter space \cite{Essig:2012yx,Essig:2011nj}. 
At Xenon 10 the non relativistic, spin-independent cross section for scattering off nucleon (N) at direct detection experiments is 
\be
\sigma_{\chi N} = \frac{|C_{\phi N}|^2 g_\chi^2 \mu_{\chi N}^2}{\pi m_\phi^4} ~,~
\ee
where $\mu_{\chi N}$ is the DM-nucleon reduced mass and the effective $\phi$ coupling 
to nucleons can be written \cite{Shifman:1978zn}
\be
C_{\phi N} =  \frac{\sqrt{2} \sin\theta \,  m_N}{v}  \left(  \frac{6}{27} \,  f^{(N)}_{TG}  +    \sum_{q = u,d,s} \!\!  f^{(N)}_{Tq}   \right),~~
\ee
where  $ f^{(N)}_{TG}, f^{(N)}_{Tq} \sim {\cal O}(0.1)$ are determined by averaging the entries in the Appendix of \cite{DelNobile:2013sia}. In Fig. 2 we present the direct detection limits
for the invisibly decaying case $m_\phi > 2m_\chi$  in terms of the parameter combination responsible for thermal freezeout. 
\be
\sigma_{\chi N} \sim      0.1   \frac{ m_N^2 }{m_e^2}       \frac{ \mu_{\chi N}^2}{\pi  m_\chi^4}      \times    (g_\chi g_e)^2\left( \frac{m_\chi}{m_\phi} \right)^{\!4}           ~,~
\ee
which, for a fixed cross section yields an estimated      
\be
 \kappa_e = (g_\chi g_e)^2\left( \frac{m_\chi}{m_\phi} \right)^{\!4} \sim  10^{-15} \left(  \frac{10^{-38} \, \cm^2}{\sigma_{\chi N}}   \right)  
 \left(  \frac{    m_\chi }{   \GeV   }   \right)^{2}\! \!.~~~~~~~~
\ee
For each $m_\chi$ there is a unique value of $\kappa_e$ constrained by direct detection, so these curves are invariant under different assumptions about the individual ratios that constitute this variable. Using this notation in Figs. 2  we present direct detection limits from LUX \cite{Akerib:2013tjd} Super-CDMS \cite{Agnese:2014aze}, and new results from CRESST-II \cite{Angloher:2015ewa}, which combine to rule out the thermal relic curve. We also show projections for upcoming and planned experiments NEWS \cite{Gerbier:2014jwa} and Super-CDMS SNOLAB \cite{Cushman:2013zza}, and LZ  \cite{Cushman:2013zza}


\section{Visibly Decaying Mediator $(m_\phi < 2m_\chi)$}
\label{sec:visible}


\begin{figure*}[t] 
\hspace{-0.4cm}
\includegraphics[width=8.5cm]{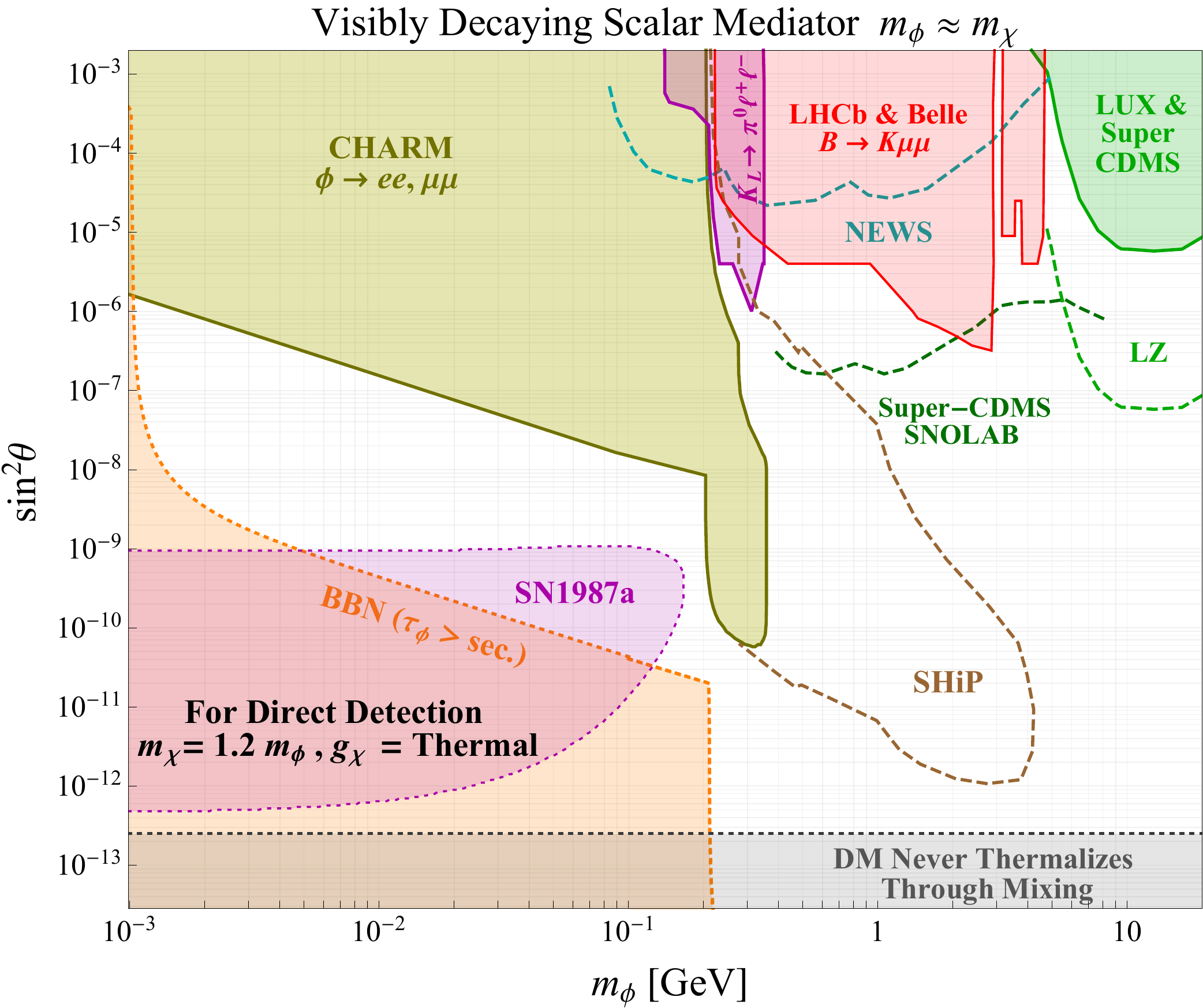} ~~
\includegraphics[width=8.5cm]{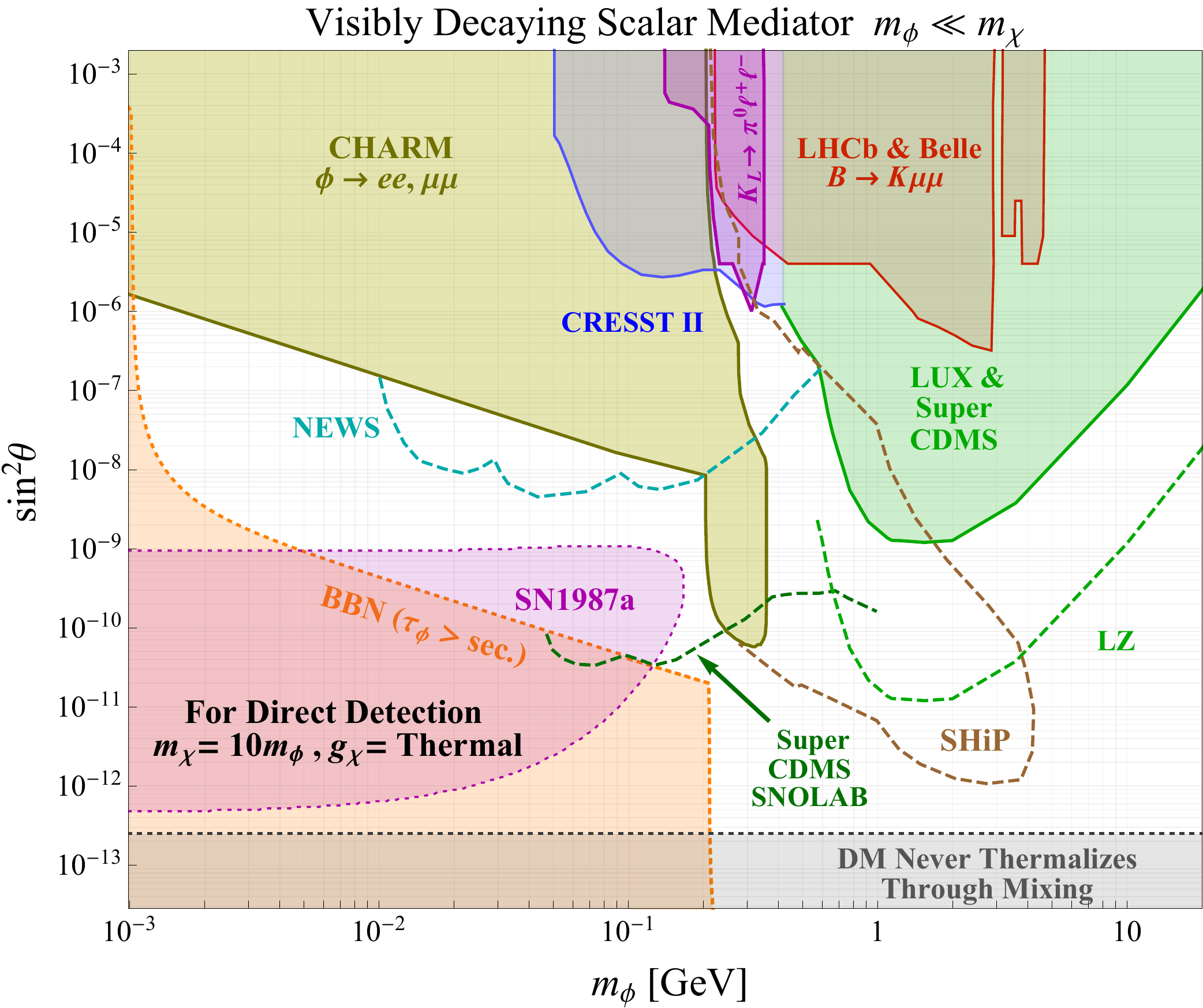} ~~
\includegraphics[width=8.5cm]{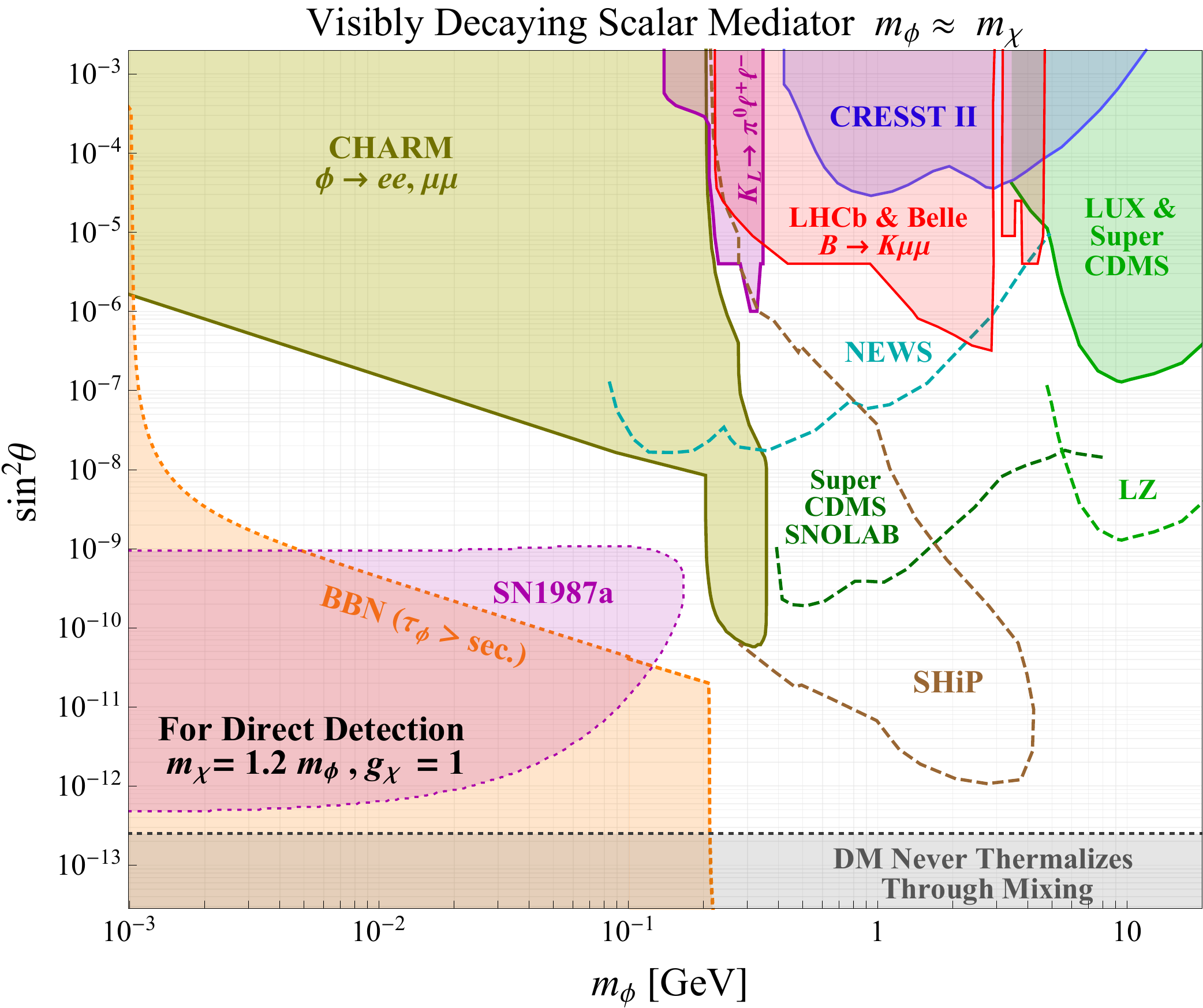}  ~~
\includegraphics[width=8.5cm]{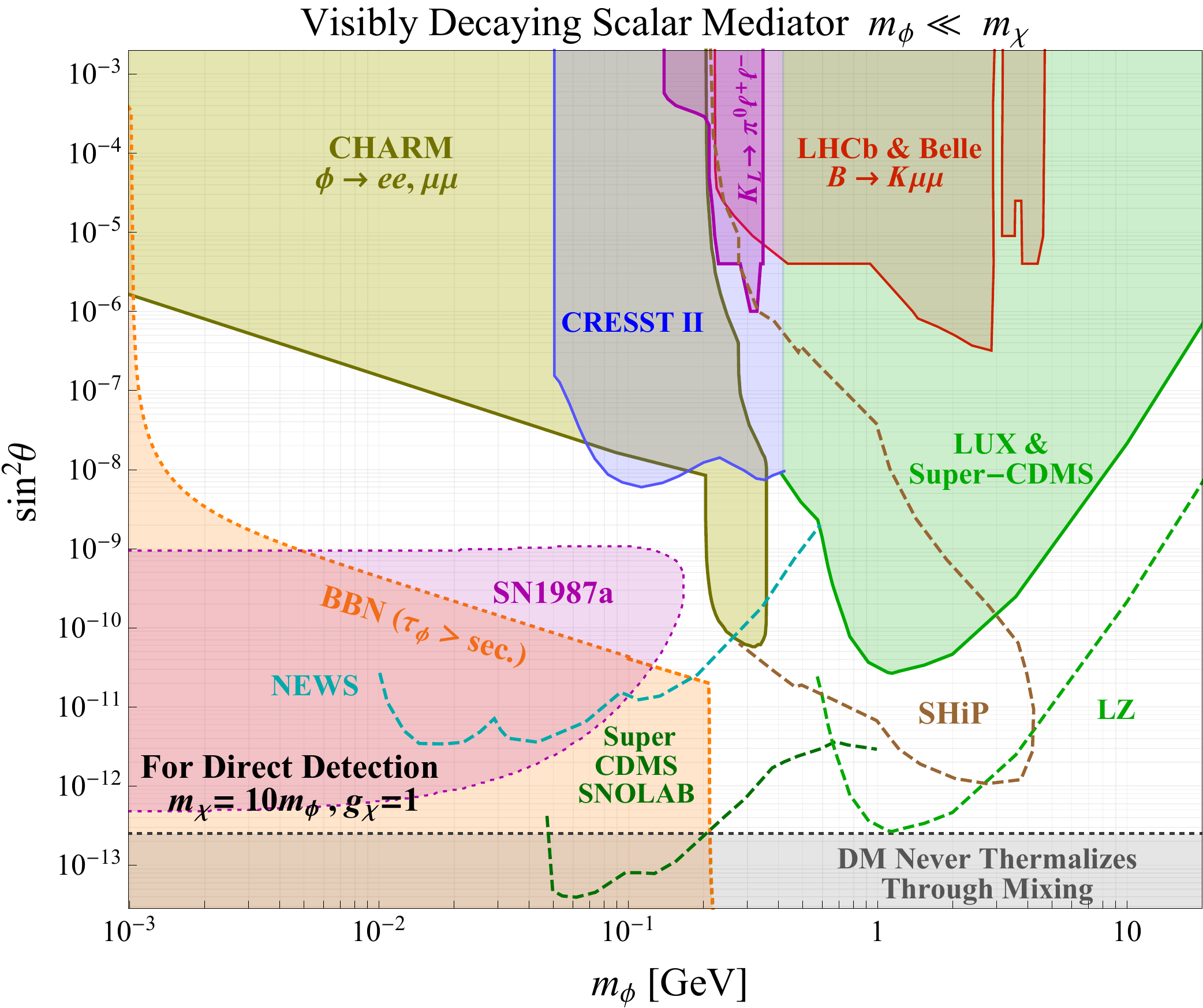}  ~~
  \caption{ {\bf } Existing constraints on the mediator-Higgs mixing in the visibly decaying $\phi \to {\rm SM ~SM}$ regime.
   {\bf Top row:} The DM is a particle-antiparticle symmetric thermal relic whose abundance is set by $t$-channel $\chi \chi \to \phi \phi$ annihilation, which 
determines the requisite $g_\chi$ coupling for a given DM mass point. Note that most of the parameter space is covered by direct searches for the mediator 
decaying into SM particles, so except for direct detection, the plots do not require any assumption about the DM provided that the mediator decays visibly. For direct detection, we show two different regimes: $m_\chi \approx m_\phi$  (but with a slightly lighter mediator) which is the least constrained regime, and  $m_\chi = 10 \, m_\phi$; for $m_\chi /m_\phi > 10$, the DM is no longer light in this parameter space, so this regime is beyond the scope of this work.  {\bf Bottom row:} Same as top row, but with $g_\chi = 1$, which corresponds to couplings larger than thermal, but still compatible with asymmetric DM, whose antiparticles have all been depleted by annihilation; these plots represent the most aggressive
bounds and projections compatible with both DM-SM equilibration and perturbative unitarity. Combined, these  four plots bracket the full parameter space of interest; smaller mass ratios than shown on the left column would invalidate the visibly decaying assumption; larger mass ratios than the right column would no longer correspond to the light DM regime; smaller  DM-mediator  couplings than the top row would overclose the universe; larger DM-mediator couplings than the bottom row would require a UV completion near
the GeV scale.  Note also that the plots in the left column show bounds from $N_{\rm eff.}$ (gray vertical dashed curves) \cite{Nollett:2013pwa} due to light DM freeze-out
after neutrino decoupling (see text); the right column  does not show this bound because for $m_\chi \sim 10 m_\phi$ the left boundary of these plots, corresponding to $m_\chi
\sim 10$ MeV, is safe from this constraint. 
    }
   \label{fig:visibleplots}
\vspace{0cm}
\end{figure*}

When the mediator is lighter than the DM, the annihilation rate for freeze out is 
decoupled from the mixing angle and there is no clear target of opportunity for 
decisive experimental coverage; the observed abundance is compatible with a wide
range of mixing angles. Nonetheless, there is still a lower bound on the SM-DM interaction
from the requirement that the DM thermalize in the early universe. 


\medskip
\noindent{\bf Thermal Equilibrium}
For sufficiently weakly coupled mediators, the DM never achieves thermal equilibrium with the visible sector radiation, so
there is no model-independent minimum annihilation rate required to avoid overclosure. Unlike in the direct annihilation
scenario (where $m_\chi < m_\phi$), the thermalization criterion now applies to the mediator since the relic abundance is set by $\chi \chi \to \phi \phi$ annihilation, 
so the DM is always  in thermal contact with the mediator. As in the former scenario, we determine the requisite $\sin \theta$ by solving 
$n_t(T) \langle \sigma v \rangle \sim H(T)$ near  $T\sim m_t$, which yields the approximate requirement 
\be
g_\chi^2 \sin^2\!\theta \gsim \frac{   53 \pi^3 \sqrt{g_*(m_t)}   \, m_t }{  \zeta(3) \, m_{Pl}           }  \approx  2 \times 10^{-13}~,~~
\ee
which defines the gray shaded region in Fig. 3. Note that this bound is insensitive to the DM and mediator masses below the electroweak scale.
Here we have taken $n_t \propto m_t^3$. For a more thorough treatment of thermalization, see the discussion in Appendix C.


\medskip
\noindent{\bf Beam-Dumps}
The CHARM Collaboration placed direct limits on light axion-like particles using a 400 GeV proton beam impinging on a copper target \cite{Bergsma:1985qz}.  In Fig. 4 we show the constraint computed in
\cite{Clarke:2013aya} (see also \cite{Schmidt-Hoberg:2013hba,Alekhin:2015byh} for similar results) in the yellow shaded region. We also show projections for an analogous
beam dump search using the SHiP facility at CERN (dashed brown contour) \cite{Alekhin:2015byh}.

\medskip


\noindent{\bf Visible Meson Decays}
A visibly decaying scalar mediator can contribute to the process $B^+ \to K^+ \mu^+\mu^-$, which is 
tightly constrained by LHCb and Belle  measurements of the branching ratio. In Fig. 3 we show the constraint
on the $\phi-h$ mixing angle as a function of light mediator mass as computed in \cite{Schmidt-Hoberg:2013hba,Clarke:2013aya}. The gaps in the coverage near $m_\phi\sim 2-3 \, \GeV$ are regions vetoed by the search criteria in \cite{Aaij:2012vr,Wei:2009zv}.
In the same parameter space, we also show bounds from measurements of the $K_L \to \pi^0 \ell^+\ell^-$ branching ratio \cite{AlaviHarati:2000hs,AlaviHarati:2003mr} as
extracted in \cite{Alekhin:2015byh} for light scalars using the analysis to bound pseduoscalar mediators in \cite{Dolan:2014ska}.


\medskip
\noindent{\bf BBN}
A sufficiently light $(m_\phi \lsim 10$ MeV), weakly coupled scalar particle with a thermal number density can decay appreciably during Big Bang Nucleosynthesis (BBN) and spoil the successful 
predictions of light element yields accumulated in the early universe. To leading order, the partial widths for SM decay channels are \cite{Djouadi:2005gi}
\be
 \Gamma(\phi \to f \bar f) &=&    s^2 \theta \,  \frac{ G_F m_f^2 m_\phi }{4 \sqrt{2}  }  \biggl(1-\frac{4m_\chi^2}{m_\phi^2} \biggr)^{\!3/2} ,~~~ \\ 
\Gamma(\phi \to \gamma\gamma )  &=&  \! s^2 \theta \frac{   G_F \, \alpha^2   m_\phi^3  }{128 \sqrt{2} \pi^3 }     \biggl|     \sum_f  N_c Q_f^2  A_{1/2}(\tau_f)  \!  + \!  A_1(\tau)   \biggr|^2 \!\!  ,~~~~~~~ \\ 
\Gamma(\phi \to gg ) &=&s^2 \theta \, \frac{  \, G_F \, \alpha^2_s(m_\phi)   m_\phi^3  }{36 \sqrt{2}\pi^3 }      \biggl|    \frac{3}{4}  \sum_q  A_{1/2}(\tau_q)      \biggr|^2 ,  ~~~~~
\ee
where $\tau_i \equiv m_\phi^2/4m_i^2$, $N_c$ is the number of colors for a given fermion species, $Q_f$ is  its electromagnetic charge, $s^2\theta \equiv \sin^2\!\theta$
and 
\be
&&A_{1/2}(\tau) = 2 [\tau + (\tau - 1)f(\tau)] \tau^{-2}  ~,~  \\
&& A_{1}(\tau) = -[2 \tau^2 + 3 \tau +3 (2\tau-1) f(\tau)] \tau^{-2}  ~,~   
\ee
where the function $f(\tau)$ satisfies 
\be
&& f(\tau)  = \begin{cases}~~~\arcsin^2\sqrt{\tau} & \tau \le 1 \\ 
-\frac{1}{4} \left[                    
\log      \frac{            1 + \sqrt{1-\tau^{-1}     }        }{     1 - \sqrt{1-\tau^{-1}  }       }
 - i \pi \right]^2 & \tau> 1 ~~, \end{cases} 
\ee
and only kinematically accessible decay channels are included for a given $m_\phi$; in our numerical 
results we only include the decay to gluons for $m_\phi >  2 m_{\pi}$.  Below the QCD confinement scale, 
we include hadronic decays following the prescription in Eq.~(\ref{eq:hadronic-phi-coupling}).
While a thorough treatment of the BBN constraint 
is beyond the scope of this paper, we safely demand  that the lifetime satisfy  $\Gamma_\phi^{-1} \lsim 1 \sec$, and 
show this boundary as the orange region in Fig. 3, which only covers thermal parameter space 
for masses where $\phi$ decays to $e^+e^-$ and $\gamma \gamma$. 

Since DM near the lower end of our mass range ($\sim$ few MeV) freezes out during the epoch of BBN, in
principle there is an additional constraint from the observed light element yields from this process as well. However,
near freeze out, the DM number density is approximately  
\be
n_{\chi} \sim  \frac{\Omega_\chi}{\Omega_b} \frac{m_{p}  }{m_\chi}  n_b  =  \frac{\Omega_\chi}{\Omega_b} \frac{m_{p}  }{m_\chi}  \eta n_\gamma  ~,~~
\ee
where $\eta \equiv n_b/n_\gamma \sim 10^{-9}$ is the baryon to photon ratio, so even on the lower end of our mass range, $n_\chi \ll n_\gamma$ during BBN. Thus, DM annihilation
products predominantly thermalize with SM photons before affecting the yields of any nuclear species during BBN. 


\medskip
\noindent{\bf Supernovae}
A light, weakly coupled scalar mediator can be produced on shell during a supernova (SN) explosion and significantly contribute to its energy loss, thereby shortening
 the duration of the observable neutrino pulse emitted during core collapse. The most significant such constraint arises from SN1987a which has been used to constrain the parameter space for axions and axion-like particles \cite{Turner:1987by,Frieman:1987ui,Burrows:1988ah,Essig:2010gu} whose SN production is dominated by radiative production off nucleons. 
 Following the prescription in  \cite{Ishizuka:1989ts}, the energy loss rate per unit volume due to $\phi$ production is   
 \be
 Q_\phi &\sim&  C^2_{\phi N}  \frac{11}{(15\pi)^3} \left(\frac{T}{m_\pi}\right)^4   p_F^5 \, G_\phi  \! \left(\frac{m_\pi}{p_F}\right)  \xi(T, m_\phi) ,~~~~~~~~ 
 \ee
 where $p_F \sim 200$ MeV is the SN Fermi momentum,  the function $G_\phi$ satisfies \cite{Ishizuka:1989ts}
 \be 
 G_\phi(u)&=& 1 - \frac{5}{2}u^2 - \frac{35}{22}u^4 + \frac{5}{64} ( 28 u^3 + 5u^5) \tan^{-1}\left(\frac{2}{u}\right) ~~~~ \nonumber  \\
  && +  \frac{5}{64} \frac{\sqrt{2} u^6}{\sqrt{2 + u^2}} \tan^{-1} \left(  \frac{2 \sqrt{2 (u^2 + 2)}     }{  u^2   }   \right)~,~~
 \ee
 and we have appended the additional factor 
 \be
 \xi(T, m_\phi) \equiv \frac{  \int_{m_\phi}^\infty   dx \frac{  x \sqrt{x^2 - m_\phi^2} }{   e^{x/T}  - 1 }    }{
  \int_0^\infty   dx \frac{  x^2}{   e^{x/T}  - 1 }   
}, 
 \ee
 to approximately account for the finite scalar mass effects, which were not included in \cite{Ishizuka:1989ts}.  
To extract an approximate bound from SN1987a, we demand that the total energy-loss rate satisfy $P_{\rm esc.} Q_\phi V_{SN} \lsim10^{53} \,  {\rm erg/s}$ where $V_{\rm SN} = (4 \pi/3) R_{\rm SN}^3$ is the SN volume, $R_{\rm SN} = 10$ km is its radius, and 
\be
P_{\rm esc.} = \exp(-R_{\rm SN} / \gamma c\tau_\phi) \exp(  - R_{\rm SN} / \lambda_\phi) ~,~
\ee
is the probability that $\phi$ escapes
without decaying or being reabsorbed within the supernova interior,
where $\gamma \tau_\phi$ is the average boosted lifetime and $\lambda_\phi$ its the mean free path. Following
 \cite{Turner:1987by}, we estimate the latter using detailed balance $\lambda_\phi^{-1} \sim Q_\phi/ \rho_\phi$ where $\rho_\phi$ is the scalar energy density estimated in the equilibrium limit for $T_{\rm SN} \sim 30$ MeV.  
 
 In Fig. 4 we 
show the excluded region shaded in purple and labeled SN1987a. Since this estimate is only valid at the 
order of magnitude level, the region is plotted with a dotted boundary; given the exponential sensitivity to some aspects
of this treatment, a dedicated study is required to extract a more rigorous bound, but is beyond the scope of this work.


\medskip
\noindent\noindent{\bf Direct Detection}
For a visibly decaying mediator, the physics of direct detection is identical to that
of the invisible case considered above.  However, since we are now interested in the $\sin^2\theta$ vs. $m_\phi$ parameter 
space, we have to make an assumption about the DM mass in order to plot the direct detection bound on the same plane in Fig. 4;
all other constraints in this parameter space depend only on the mixing and the mediator mass. Nonetheless, we can still comprehensively capture the parameter space
compatible with light, thermal DM  ($<$ few GeV) by looking at the two extreme choices of mass ratios: $m_\chi = 1.2 m_\phi$ (Fig. 3, left column) and $m_\chi = 10 m_\phi$ (Fig. 3, right column) for much smaller ratios, the mediator starts to decay visibly and for much larger ratios the DM is no longer ``light". In addition to specifying the DM-mediator mass ratio, we 
must also choose $g_\chi$ to extract a limit on $\sin\theta$. For the top row of Fig. 3, our choice (for a fixed DM-mediator mass ratio) we choose $g_\chi$ to yield the observed 
relic abundance through symmetric thermal freeze out via $\chi\bar\chi\to\phi\phi$; this is the most conservative choice as it corresponds to the smallest coupling consistent 
with thermal relic DM and correspondingly yields the weakest direct detection constraints and future projections. For the lower row in Fig. 4 we consider the opposite regime $g_\chi =1$ which is near the unitarity boundary and (if $\chi$ constitutes all the DM) is compatible with asymmetric DM scenario in which the larger-than-thermal annihilation rate 
efficiently depletes most of the dark antiparticles produced in thermal equilibrium with the visible sector


\section{Visibly Decaying Mediator, \\ Compressed Regime $( m_\chi < m_\phi < 2m_\chi)$}
\label{sec:compressed}

In the compressed region of parameter space where $m_\chi < m_\phi < 2m_\chi$, DM annihilation
proceeds through an off-shell $\phi$ via $\chi \chi \to \phi*\to ff$, which depends on $\kappa_e$ as described in 
Sec. \ref{sec:thermalrelic}. 
Furthermore, since $m_\chi /m_\phi \sim 1$, we have
\be
\kappa_e = (g_\chi g_e)^2 \left(\frac{m_\chi}{m_\phi} \right)^4  
&\approx & g_\chi^2  \left(\frac{m_e}{v} \sin \theta \right)^2  \nonumber \\
&=& 4.3 \times 10^{-12} g_\chi^2 \sin^2 \theta~~,~~~
\ee
but from Eq.~(\ref{eq:kappa-freezeout}), thermal freeze out in the light DM
mass range requires $\kappa_e \sim 10^{-13} (m_\chi/  \rm MeV)^2$, so the requisite 
mixing angle must be $\sim {\cal O}(1)$ for a sufficiently large annihilation rate.  
However, unlike the regime considered in Sec. \ref{sec:invisible}, for the  $m_\phi < 2 m_\chi$ mass range, 
the mediator only decays to SM fermions, so all the constraints from  Sec.\ref{sec:visible} are applicable. 
From  Fig. \ref{fig:visibleplots}, it is clear that, so long as $\phi$ decays visibly, the region $\sin^2 \theta \gtrsim 10^{-5}$ is 
excluded over
the full MeV-GeV mass range. Thus, we conclude that this compressed region of parameter space is  ruled out for thermal DM. 
 

\section{\bf  Dark Matter Candidate Variations}\label{sec:variations}
We now consider simple variations away from the particle-antiparticle symmetric  
Dirac fermion DM candidate from the benchmark model in Eqs.~(\ref{eq:phiDMlag}) and (\ref{eq:phiSMlag}).
In short, we find that for scenarios involving Majorana DM, scalar DM, asymmetric DM,
or fermion DM with parity odd couplings, the analogous parameter space is either qualitatively unchanged or more
severely constrained relative to the symmetric Dirac scenario with parity even interactions. The following 
analysis, therefore, justifies our choice of 
benchmark model as sufficiently representative of the remaining viable parameter
space.


\medskip
\noindent  {\bf Parity Odd Couplings} 
In the Lagrangian for our benchmark scenario, Eq.~(\ref{eq:phiDMlag}), 
allows a  $\phi \bar \psi \gamma^5 \psi$ term. This interaction gives rise to $s$ wave annihilation for both 
$\chi \bar \chi \to f\bar f $ and  $\chi \bar \chi \to \phi \phi$ channels, which is robustly ruled out for  
thermal relic DM below $\lsim 10$ GeV (see Sec. \ref{sec:generic}), so at minimum, we require
 $g^\prime_{\chi} \ll g_{\chi}$. Although in principle, we could keep this subdominant in our analysis, 
its presence would not significantly change the qualitative character of any plots; it would only introduce 
percent level corrections for couplings that evade CMB bounds, so we have not lost
any essential generality by setting $g_\chi^\prime = 0$ in the analysis presented above.


\medskip
\noindent  {\bf Majorana DM} 
The simplest variation on our benchmark scenario involves exchanging our Dirac DM 
with a Majorana fermion. However, most of the constraints we encounter below involve
accelerator production, meson decays, or direct detection; none of which differs substantially 
with this modification. Thus, the constraints and relic density projections will differ only by order
one amounts.


\medskip
\noindent  {\bf Scalar Symmetric DM} 
If the  DM, itself, is a stable scalar $\varphi$ that couples to the Higgs-portal mediator $\phi$,  the relic abundance can be achieved through either direct annihilation $\varphi \varphi \to \phi^* \to f \bar f$  ($m_\varphi > m_\phi$) or
t-channel $\varphi \varphi \to \phi \phi$ ($m_\varphi > m_\phi$) annihilation. In both cases, the leading
 annihilation rates will be $s$-wave (velocity independent) and, therefore,
ruled out by CMB power-injection limits for $\varphi$ masses below 10 GeV \cite{Ade:2015xua}.



\medskip
\noindent {\bf Asymmetric DM}   
If the dark matter abundance is set by a primordial asymmetry \cite{Kaplan:1991ah,Nussinov:1985xr,Barr:1991qn,Barr:1990ca,Kaplan:2009ag} and the dark matter achieves thermal equilibrium with the SM, the
annihilation rate must be larger than the nominal freeze out value \cite{Graesser:2011wi,Lin:2011gj}. However, all other constraints on its direct production and detection
remain identical; since antiparticles are exponentially depleted below $T\lsim m_{\rm DM}$, the indirect detection signatures are generically suppressed at late times, so asymmetric DM can 
be compatible with the CMB. 
Thus, the results in this paper apply fully well to this variation, but the parameter space compatible with the observed abundance (for the direct annihilation topology $m_\chi < m_\phi$) is above the solid black curve in Fig. 2 (see discussion below), so this is also ruled out; for particle-antiparticle symmetric scalar DM, which has an $s$-wave annihilation rate, the relic density contour will shift slightly 
lower in the parameter space by a an ${\cal O}(1)$ amount, but  everything else remains quantitatively similar.  \medskip


\medskip 
\noindent {\bf Inelastically Coupled DM} 
In principle, an extended dark sector can couple inelastically to the mediator $\phi$, which can sharply suppress 
direct detection limits. A full treatment of this scenario is beyond the scope of this work, but we note that a minimal DM model
can not easily ensure that the  singlet  scalar couple predominantly inelastically to either SM or DM fields. The simplest such mechanism in the case of a vector mediator
involves a pseudo-Dirac splitting of Weyl fermions with both Dirac and Majorana masses \cite{TuckerSmith:2001hy}, which yields a predominantly inelastic coupling in the mass eigenbasis
and yields distinctive direct detection and collider phenomenogloy \cite{Bai:2011jg,Izaguirre:2015zva}.
However, for a scalar mediator, the analogous procedure yields comparable elastic and inelastic contributions, so the scenario is qualitatively similar to the benchmark
model we consider throughout this paper; all of the same bounds will apply, but there will be order one variations from the existence of additional couplings.


\bigskip
\bigskip
\section{Concluding Remarks}
\label{sec:conclusion}

In this paper we have mapped out parameter space for light ($<$ few GeV) thermal DM coupled 
to a light scalar that mixes with the SM Higgs boson. Although we have focused primarily on a benchmark model with a Dirac fermion DM candidate, this scenario suffices to capture most of the essential physics, so our central conclusions apply to the simplest DM variations (e.g. asymmetric  DM or scalar DM with the same mediator). 

In the more predictive, heavier-mediator regime ($m_\phi > m_\chi$) we find that thermal DM is now conservatively ruled out by a combination of 
 rare meson decay, direct detection, and off-shell Higgs width measurements. This conclusion holds regardless of the DM/mediator mass ratio so long as the latter is heavier. We also find that benchmark future low-threshold direct detection experiments (NEWS, Super-CDMS SNOLAB,  LZ and others) can improve coverage for the Higgs portal interaction, but (since the thermal relic parameter space is ruled out) this improvement would only test a subdominant interaction in the dark sector -- additional interactions would be required to avoid overclosure.
 
  Relaxing the requirement that the light scalar mediator mix exclusively with the SM Higgs doublet, it may be possible to engineer a viable direct-annihilation
 scenario in a type-III two Higgs doublet model in which one double couples to quarks and the other couples to leptons. If the light scalar mixes predominantly with the leptonic
 doublet, it may be possible to relax the constraints from rare meson decays, but we leave this investigation for future work.
 
For the lighter mediator regime ($m_\phi > m_\chi$) in which DM annihilation proceeds through the ``secluded" channel $\chi \chi \to \phi \phi$, there is no clear relic density target since the rate is independent of the Higgs-mediator mixing; thermal freeze out is compatible with a wide range of SM-mediator couplings. However, the mixing angle is bounded from below by the requirement that the DM be produced thermally in the early universe and much of the parameter space (with notable gaps) is covered  by a combination beam dump, supernovae, and BBN constraints below mediator masses of $\sim$ 100 MeV. Above this range, there are additional constraints from direct detection experiments, which are poised to improve with the next generation of efforts, but the amount of new coverage will depend greatly on the DM/mediator mass ratio. 

Finally, for the ``compressed" regime ($m_\chi < m_\phi < 2 m_\chi$) in which DM and mediator masses are nearly degenerate, 
DM annihilation proceeds via $\chi \chi \to \phi^* \to f f$, so there is a thermal freeze out requirement on the Higgs-$\phi$ mixing angle. However, in this 
mass range, the mediator decays to visible SM states, so all the bounds from the $m_\phi > m_\chi$ mass range apply equally well and rule out the 
range of Higgs-$\phi$ mixings required for thermal freeze out.   
%

\medskip  


\noindent{\it Acknowledgments}: We thank Wolfgang Altmanshoffer, Jackson Clarke, Roni Harnik, Eder Izaguirre, Maxim Pospelov, David Pinnner, Kai Schmidt-Hoberg, Philip Schuster, Brian Shuve, Flip Tanedo, Jesse Thaler, and Natalia Toro for many helpful conversations. We also thank the University of Victoria for hospitality while this work was being completed. Fermilab is operated by Fermi Research Alliance, LLC, under Contract No. DE- AC02-07CH11359 with the US Department of Energy. 
\medskip


\medskip

\bibliography{scalarprl}

\begin{thebibliography}{80}%
\makeatletter
\providecommand \@ifxundefined [1]{%
 \@ifx{#1\undefined}
}%
\providecommand \@ifnum [1]{%
 \ifnum #1\expandafter \@firstoftwo
 \else \expandafter \@secondoftwo
 \fi
}%
\providecommand \@ifx [1]{%
 \ifx #1\expandafter \@firstoftwo
 \else \expandafter \@secondoftwo
 \fi
}%
\providecommand \natexlab [1]{#1}%
\providecommand \enquote  [1]{``#1''}%
\providecommand \bibnamefont  [1]{#1}%
\providecommand \bibfnamefont [1]{#1}%
\providecommand \citenamefont [1]{#1}%
\providecommand \href@noop [0]{\@secondoftwo}%
\providecommand \href [0]{\begingroup \@sanitize@url \@href}%
\providecommand \@href[1]{\@@startlink{#1}\@@href}%
\providecommand \@@href[1]{\endgroup#1\@@endlink}%
\providecommand \@sanitize@url [0]{\catcode `\\12\catcode `\$12\catcode
  `\&12\catcode `\#12\catcode `\^12\catcode `\_12\catcode `\%12\relax}%
\providecommand \@@startlink[1]{}%
\providecommand \@@endlink[0]{}%
\providecommand \url  [0]{\begingroup\@sanitize@url \@url }%
\providecommand \@url [1]{\endgroup\@href {#1}{\urlprefix }}%
\providecommand \urlprefix  [0]{URL }%
\providecommand \Eprint [0]{\href }%
\providecommand \doibase [0]{http://dx.doi.org/}%
\providecommand \selectlanguage [0]{\@gobble}%
\providecommand \bibinfo  [0]{\@secondoftwo}%
\providecommand \bibfield  [0]{\@secondoftwo}%
\providecommand \translation [1]{[#1]}%
\providecommand \BibitemOpen [0]{}%
\providecommand \bibitemStop [0]{}%
\providecommand \bibitemNoStop [0]{.\EOS\space}%
\providecommand \EOS [0]{\spacefactor3000\relax}%
\providecommand \BibitemShut  [1]{\csname bibitem#1\endcsname}%
\let\auto@bib@innerbib\@empty
\bibitem [{\citenamefont {Lee}\ and\ \citenamefont
  {Weinberg}(1977)}]{Lee:1977ua}%
  \BibitemOpen
  \bibfield  {author} {\bibinfo {author} {\bibfnamefont {B.~W.}\ \bibnamefont
  {Lee}}\ and\ \bibinfo {author} {\bibfnamefont {S.}~\bibnamefont {Weinberg}},\
  }\href {\doibase 10.1103/PhysRevLett.39.165} {\bibfield  {journal} {\bibinfo
  {journal} {Phys.Rev.Lett.}\ }\textbf {\bibinfo {volume} {39}},\ \bibinfo
  {pages} {165} (\bibinfo {year} {1977})}\BibitemShut {NoStop}%
\bibitem [{\citenamefont {Shakya}(2015)}]{Shakya:2015xnx}%
  \BibitemOpen
  \bibfield  {author} {\bibinfo {author} {\bibfnamefont {B.}~\bibnamefont
  {Shakya}},\ }\href@noop {} {\  (\bibinfo {year} {2015})},\ \Eprint
  {http://arxiv.org/abs/1512.02751} {arXiv:1512.02751 [hep-ph]} \BibitemShut
  {NoStop}%
\bibitem [{\citenamefont {Izaguirre}\ \emph
  {et~al.}(2015{\natexlab{a}})\citenamefont {Izaguirre}, \citenamefont
  {Krnjaic}, \citenamefont {Schuster},\ and\ \citenamefont
  {Toro}}]{Izaguirre:2015yja}%
  \BibitemOpen
  \bibfield  {author} {\bibinfo {author} {\bibfnamefont {E.}~\bibnamefont
  {Izaguirre}}, \bibinfo {author} {\bibfnamefont {G.}~\bibnamefont {Krnjaic}},
  \bibinfo {author} {\bibfnamefont {P.}~\bibnamefont {Schuster}}, \ and\
  \bibinfo {author} {\bibfnamefont {N.}~\bibnamefont {Toro}},\ }\href {\doibase
  10.1103/PhysRevLett.115.251301} {\bibfield  {journal} {\bibinfo  {journal}
  {Phys. Rev. Lett.}\ }\textbf {\bibinfo {volume} {115}},\ \bibinfo {pages}
  {251301} (\bibinfo {year} {2015}{\natexlab{a}})},\ \Eprint
  {http://arxiv.org/abs/1505.00011} {arXiv:1505.00011 [hep-ph]} \BibitemShut
  {NoStop}%
\bibitem [{\citenamefont {Essig}\ \emph {et~al.}(2013)\citenamefont {Essig},
  \citenamefont {Jaros}, \citenamefont {Wester}, \citenamefont {Adrian},
  \citenamefont {Andreas} \emph {et~al.}}]{Essig:2013lka}%
  \BibitemOpen
  \bibfield  {author} {\bibinfo {author} {\bibfnamefont {R.}~\bibnamefont
  {Essig}}, \bibinfo {author} {\bibfnamefont {J.~A.}\ \bibnamefont {Jaros}},
  \bibinfo {author} {\bibfnamefont {W.}~\bibnamefont {Wester}}, \bibinfo
  {author} {\bibfnamefont {P.~H.}\ \bibnamefont {Adrian}}, \bibinfo {author}
  {\bibfnamefont {S.}~\bibnamefont {Andreas}},  \emph {et~al.},\ }\href@noop {}
  {\  (\bibinfo {year} {2013})},\ \Eprint {http://arxiv.org/abs/1311.0029}
  {arXiv:1311.0029 [hep-ph]} \BibitemShut {NoStop}%
\bibitem [{\citenamefont {Essig}\ \emph {et~al.}(2011)\citenamefont {Essig},
  \citenamefont {Schuster}, \citenamefont {Toro},\ and\ \citenamefont
  {Wojtsekhowski}}]{Essig:2010xa}%
  \BibitemOpen
  \bibfield  {author} {\bibinfo {author} {\bibfnamefont {R.}~\bibnamefont
  {Essig}}, \bibinfo {author} {\bibfnamefont {P.}~\bibnamefont {Schuster}},
  \bibinfo {author} {\bibfnamefont {N.}~\bibnamefont {Toro}}, \ and\ \bibinfo
  {author} {\bibfnamefont {B.}~\bibnamefont {Wojtsekhowski}},\ }\href {\doibase
  10.1007/JHEP02(2011)009} {\bibfield  {journal} {\bibinfo  {journal} {JHEP}\
  }\textbf {\bibinfo {volume} {1102}},\ \bibinfo {pages} {009} (\bibinfo {year}
  {2011})},\ \Eprint {http://arxiv.org/abs/1001.2557} {arXiv:1001.2557
  [hep-ph]} \BibitemShut {NoStop}%
\bibitem [{\citenamefont {Merkel}\ \emph {et~al.}(2011)\citenamefont {Merkel}
  \emph {et~al.}}]{Merkel:2011ze}%
  \BibitemOpen
  \bibfield  {author} {\bibinfo {author} {\bibfnamefont {H.}~\bibnamefont
  {Merkel}} \emph {et~al.} (\bibinfo {collaboration} {A1 Collaboration}),\
  }\href {\doibase 10.1103/PhysRevLett.106.251802} {\bibfield  {journal}
  {\bibinfo  {journal} {Phys.Rev.Lett.}\ }\textbf {\bibinfo {volume} {106}},\
  \bibinfo {pages} {251802} (\bibinfo {year} {2011})},\ \Eprint
  {http://arxiv.org/abs/1101.4091} {arXiv:1101.4091 [nucl-ex]} \BibitemShut
  {NoStop}%
\bibitem [{\citenamefont {Abrahamyan}\ \emph {et~al.}(2011)\citenamefont
  {Abrahamyan} \emph {et~al.}}]{Abrahamyan:2011gv}%
  \BibitemOpen
  \bibfield  {author} {\bibinfo {author} {\bibfnamefont {S.}~\bibnamefont
  {Abrahamyan}} \emph {et~al.} (\bibinfo {collaboration} {APEX
  Collaboration}),\ }\href {\doibase 10.1103/PhysRevLett.107.191804} {\bibfield
   {journal} {\bibinfo  {journal} {Phys.Rev.Lett.}\ }\textbf {\bibinfo {volume}
  {107}},\ \bibinfo {pages} {191804} (\bibinfo {year} {2011})},\ \Eprint
  {http://arxiv.org/abs/1108.2750} {arXiv:1108.2750 [hep-ex]} \BibitemShut
  {NoStop}%
\bibitem [{\citenamefont {Izaguirre}\ \emph {et~al.}(2013)\citenamefont
  {Izaguirre}, \citenamefont {Krnjaic}, \citenamefont {Schuster},\ and\
  \citenamefont {Toro}}]{Izaguirre:2013uxa}%
  \BibitemOpen
  \bibfield  {author} {\bibinfo {author} {\bibfnamefont {E.}~\bibnamefont
  {Izaguirre}}, \bibinfo {author} {\bibfnamefont {G.}~\bibnamefont {Krnjaic}},
  \bibinfo {author} {\bibfnamefont {P.}~\bibnamefont {Schuster}}, \ and\
  \bibinfo {author} {\bibfnamefont {N.}~\bibnamefont {Toro}},\ }\href {\doibase
  10.1103/PhysRevD.88.114015} {\bibfield  {journal} {\bibinfo  {journal} {Phys.
  Rev.}\ }\textbf {\bibinfo {volume} {D88}},\ \bibinfo {pages} {114015}
  (\bibinfo {year} {2013})},\ \Eprint {http://arxiv.org/abs/1307.6554}
  {arXiv:1307.6554 [hep-ph]} \BibitemShut {NoStop}%
\bibitem [{\citenamefont {Dharmapalan}\ \emph {et~al.}(2012)\citenamefont
  {Dharmapalan} \emph {et~al.}}]{Dharmapalan:2012xp}%
  \BibitemOpen
  \bibfield  {author} {\bibinfo {author} {\bibfnamefont {R.}~\bibnamefont
  {Dharmapalan}} \emph {et~al.} (\bibinfo {collaboration} {MiniBooNE}),\
  }\href@noop {} {\  (\bibinfo {year} {2012})},\ \Eprint
  {http://arxiv.org/abs/1211.2258} {arXiv:1211.2258 [hep-ex]} \BibitemShut
  {NoStop}%
\bibitem [{\citenamefont {Izaguirre}\ \emph
  {et~al.}(2015{\natexlab{b}})\citenamefont {Izaguirre}, \citenamefont
  {Krnjaic},\ and\ \citenamefont {Pospelov}}]{Izaguirre:2015pva}%
  \BibitemOpen
  \bibfield  {author} {\bibinfo {author} {\bibfnamefont {E.}~\bibnamefont
  {Izaguirre}}, \bibinfo {author} {\bibfnamefont {G.}~\bibnamefont {Krnjaic}},
  \ and\ \bibinfo {author} {\bibfnamefont {M.}~\bibnamefont {Pospelov}},\
  }\href {\doibase 10.1103/PhysRevD.92.095014} {\bibfield  {journal} {\bibinfo
  {journal} {Phys. Rev.}\ }\textbf {\bibinfo {volume} {D92}},\ \bibinfo {pages}
  {095014} (\bibinfo {year} {2015}{\natexlab{b}})},\ \Eprint
  {http://arxiv.org/abs/1507.02681} {arXiv:1507.02681 [hep-ph]} \BibitemShut
  {NoStop}%
\bibitem [{\citenamefont {Izaguirre}\ \emph {et~al.}(2014)\citenamefont
  {Izaguirre}, \citenamefont {Krnjaic}, \citenamefont {Schuster},\ and\
  \citenamefont {Toro}}]{Izaguirre:2014bca}%
  \BibitemOpen
  \bibfield  {author} {\bibinfo {author} {\bibfnamefont {E.}~\bibnamefont
  {Izaguirre}}, \bibinfo {author} {\bibfnamefont {G.}~\bibnamefont {Krnjaic}},
  \bibinfo {author} {\bibfnamefont {P.}~\bibnamefont {Schuster}}, \ and\
  \bibinfo {author} {\bibfnamefont {N.}~\bibnamefont {Toro}},\ }\href@noop {}
  {\  (\bibinfo {year} {2014})},\ \Eprint {http://arxiv.org/abs/1411.1404}
  {arXiv:1411.1404 [hep-ph]} \BibitemShut {NoStop}%
\bibitem [{\citenamefont {Batell}\ \emph {et~al.}(2014)\citenamefont {Batell},
  \citenamefont {Essig},\ and\ \citenamefont {Surujon}}]{Batell:2014mga}%
  \BibitemOpen
  \bibfield  {author} {\bibinfo {author} {\bibfnamefont {B.}~\bibnamefont
  {Batell}}, \bibinfo {author} {\bibfnamefont {R.}~\bibnamefont {Essig}}, \
  and\ \bibinfo {author} {\bibfnamefont {Z.}~\bibnamefont {Surujon}},\ }\href
  {\doibase 10.1103/PhysRevLett.113.171802} {\bibfield  {journal} {\bibinfo
  {journal} {Phys.Rev.Lett.}\ }\textbf {\bibinfo {volume} {113}},\ \bibinfo
  {pages} {171802} (\bibinfo {year} {2014})},\ \Eprint
  {http://arxiv.org/abs/1406.2698} {arXiv:1406.2698 [hep-ph]} \BibitemShut
  {NoStop}%
\bibitem [{\citenamefont {Kahn}\ \emph {et~al.}(2014)\citenamefont {Kahn},
  \citenamefont {Krnjaic}, \citenamefont {Thaler},\ and\ \citenamefont
  {Toups}}]{Kahn:2014sra}%
  \BibitemOpen
  \bibfield  {author} {\bibinfo {author} {\bibfnamefont {Y.}~\bibnamefont
  {Kahn}}, \bibinfo {author} {\bibfnamefont {G.}~\bibnamefont {Krnjaic}},
  \bibinfo {author} {\bibfnamefont {J.}~\bibnamefont {Thaler}}, \ and\ \bibinfo
  {author} {\bibfnamefont {M.}~\bibnamefont {Toups}},\ }\href@noop {} {\
  (\bibinfo {year} {2014})},\ \Eprint {http://arxiv.org/abs/1411.1055}
  {arXiv:1411.1055 [hep-ph]} \BibitemShut {NoStop}%
\bibitem [{\citenamefont {Battaglieri}\ \emph {et~al.}(2014)\citenamefont
  {Battaglieri} \emph {et~al.}}]{Battaglieri:2014qoa}%
  \BibitemOpen
  \bibfield  {author} {\bibinfo {author} {\bibfnamefont {M.}~\bibnamefont
  {Battaglieri}} \emph {et~al.} (\bibinfo {collaboration} {BDX
  Collaboration}),\ }\href@noop {} {\  (\bibinfo {year} {2014})},\ \Eprint
  {http://arxiv.org/abs/1406.3028} {arXiv:1406.3028 [physics.ins-det]}
  \BibitemShut {NoStop}%
\bibitem [{\citenamefont {Hochberg}\ \emph {et~al.}(2015)\citenamefont
  {Hochberg}, \citenamefont {Zhao},\ and\ \citenamefont
  {Zurek}}]{Hochberg:2015pha}%
  \BibitemOpen
  \bibfield  {author} {\bibinfo {author} {\bibfnamefont {Y.}~\bibnamefont
  {Hochberg}}, \bibinfo {author} {\bibfnamefont {Y.}~\bibnamefont {Zhao}}, \
  and\ \bibinfo {author} {\bibfnamefont {K.~M.}\ \bibnamefont {Zurek}},\
  }\href@noop {} {\  (\bibinfo {year} {2015})},\ \Eprint
  {http://arxiv.org/abs/1504.07237} {arXiv:1504.07237 [hep-ph]} \BibitemShut
  {NoStop}%
\bibitem [{\citenamefont {Curtin}\ \emph {et~al.}(2015)\citenamefont {Curtin},
  \citenamefont {Essig}, \citenamefont {Gori},\ and\ \citenamefont
  {Shelton}}]{Curtin:2014cca}%
  \BibitemOpen
  \bibfield  {author} {\bibinfo {author} {\bibfnamefont {D.}~\bibnamefont
  {Curtin}}, \bibinfo {author} {\bibfnamefont {R.}~\bibnamefont {Essig}},
  \bibinfo {author} {\bibfnamefont {S.}~\bibnamefont {Gori}}, \ and\ \bibinfo
  {author} {\bibfnamefont {J.}~\bibnamefont {Shelton}},\ }\href {\doibase
  10.1007/JHEP02(2015)157} {\bibfield  {journal} {\bibinfo  {journal} {JHEP}\
  }\textbf {\bibinfo {volume} {02}},\ \bibinfo {pages} {157} (\bibinfo {year}
  {2015})},\ \Eprint {http://arxiv.org/abs/1412.0018} {arXiv:1412.0018
  [hep-ph]} \BibitemShut {NoStop}%
\bibitem [{\citenamefont {D'Agnolo}\ and\ \citenamefont
  {Ruderman}(2015)}]{D'Agnolo:2015koa}%
  \BibitemOpen
  \bibfield  {author} {\bibinfo {author} {\bibfnamefont {R.~T.}\ \bibnamefont
  {D'Agnolo}}\ and\ \bibinfo {author} {\bibfnamefont {J.~T.}\ \bibnamefont
  {Ruderman}},\ }\href {\doibase 10.1103/PhysRevLett.115.061301} {\bibfield
  {journal} {\bibinfo  {journal} {Phys. Rev. Lett.}\ }\textbf {\bibinfo
  {volume} {115}},\ \bibinfo {pages} {061301} (\bibinfo {year} {2015})},\
  \Eprint {http://arxiv.org/abs/1505.07107} {arXiv:1505.07107 [hep-ph]}
  \BibitemShut {NoStop}%
\bibitem [{\citenamefont {Burgess}\ \emph {et~al.}(2001)\citenamefont
  {Burgess}, \citenamefont {Pospelov},\ and\ \citenamefont {ter
  Veldhuis}}]{Burgess:2000yq}%
  \BibitemOpen
  \bibfield  {author} {\bibinfo {author} {\bibfnamefont {C.~P.}\ \bibnamefont
  {Burgess}}, \bibinfo {author} {\bibfnamefont {M.}~\bibnamefont {Pospelov}}, \
  and\ \bibinfo {author} {\bibfnamefont {T.}~\bibnamefont {ter Veldhuis}},\
  }\href {\doibase 10.1016/S0550-3213(01)00513-2} {\bibfield  {journal}
  {\bibinfo  {journal} {Nucl. Phys.}\ }\textbf {\bibinfo {volume} {B619}},\
  \bibinfo {pages} {709} (\bibinfo {year} {2001})},\ \Eprint
  {http://arxiv.org/abs/hep-ph/0011335} {arXiv:hep-ph/0011335 [hep-ph]}
  \BibitemShut {NoStop}%
\bibitem [{\citenamefont {Pospelov}\ and\ \citenamefont
  {Ritz}(2011)}]{Pospelov:2011yp}%
  \BibitemOpen
  \bibfield  {author} {\bibinfo {author} {\bibfnamefont {M.}~\bibnamefont
  {Pospelov}}\ and\ \bibinfo {author} {\bibfnamefont {A.}~\bibnamefont
  {Ritz}},\ }\href {\doibase 10.1103/PhysRevD.84.113001} {\bibfield  {journal}
  {\bibinfo  {journal} {Phys. Rev.}\ }\textbf {\bibinfo {volume} {D84}},\
  \bibinfo {pages} {113001} (\bibinfo {year} {2011})},\ \Eprint
  {http://arxiv.org/abs/1109.4872} {arXiv:1109.4872 [hep-ph]} \BibitemShut
  {NoStop}%
\bibitem [{\citenamefont {Bird}\ \emph {et~al.}(2006)\citenamefont {Bird},
  \citenamefont {Kowalewski},\ and\ \citenamefont {Pospelov}}]{Bird:2006jd}%
  \BibitemOpen
  \bibfield  {author} {\bibinfo {author} {\bibfnamefont {C.}~\bibnamefont
  {Bird}}, \bibinfo {author} {\bibfnamefont {R.~V.}\ \bibnamefont
  {Kowalewski}}, \ and\ \bibinfo {author} {\bibfnamefont {M.}~\bibnamefont
  {Pospelov}},\ }\href {\doibase 10.1142/S0217732306019852} {\bibfield
  {journal} {\bibinfo  {journal} {Mod. Phys. Lett.}\ }\textbf {\bibinfo
  {volume} {A21}},\ \bibinfo {pages} {457} (\bibinfo {year} {2006})},\ \Eprint
  {http://arxiv.org/abs/hep-ph/0601090} {arXiv:hep-ph/0601090 [hep-ph]}
  \BibitemShut {NoStop}%
\bibitem [{\citenamefont {Pospelov}\ \emph {et~al.}(2008)\citenamefont
  {Pospelov}, \citenamefont {Ritz},\ and\ \citenamefont
  {Voloshin}}]{Pospelov:2007mp}%
  \BibitemOpen
  \bibfield  {author} {\bibinfo {author} {\bibfnamefont {M.}~\bibnamefont
  {Pospelov}}, \bibinfo {author} {\bibfnamefont {A.}~\bibnamefont {Ritz}}, \
  and\ \bibinfo {author} {\bibfnamefont {M.~B.}\ \bibnamefont {Voloshin}},\
  }\href {\doibase 10.1016/j.physletb.2008.02.052} {\bibfield  {journal}
  {\bibinfo  {journal} {Phys. Lett.}\ }\textbf {\bibinfo {volume} {B662}},\
  \bibinfo {pages} {53} (\bibinfo {year} {2008})},\ \Eprint
  {http://arxiv.org/abs/0711.4866} {arXiv:0711.4866 [hep-ph]} \BibitemShut
  {NoStop}%
\bibitem [{\citenamefont {Bird}\ \emph {et~al.}(2004)\citenamefont {Bird},
  \citenamefont {Jackson}, \citenamefont {Kowalewski},\ and\ \citenamefont
  {Pospelov}}]{Bird:2004ts}%
  \BibitemOpen
  \bibfield  {author} {\bibinfo {author} {\bibfnamefont {C.}~\bibnamefont
  {Bird}}, \bibinfo {author} {\bibfnamefont {P.}~\bibnamefont {Jackson}},
  \bibinfo {author} {\bibfnamefont {R.~V.}\ \bibnamefont {Kowalewski}}, \ and\
  \bibinfo {author} {\bibfnamefont {M.}~\bibnamefont {Pospelov}},\ }\href
  {\doibase 10.1103/PhysRevLett.93.201803} {\bibfield  {journal} {\bibinfo
  {journal} {Phys. Rev. Lett.}\ }\textbf {\bibinfo {volume} {93}},\ \bibinfo
  {pages} {201803} (\bibinfo {year} {2004})},\ \Eprint
  {http://arxiv.org/abs/hep-ph/0401195} {arXiv:hep-ph/0401195 [hep-ph]}
  \BibitemShut {NoStop}%
\bibitem [{\citenamefont {Schmidt-Hoberg}\ \emph {et~al.}(2013)\citenamefont
  {Schmidt-Hoberg}, \citenamefont {Staub},\ and\ \citenamefont
  {Winkler}}]{Schmidt-Hoberg:2013hba}%
  \BibitemOpen
  \bibfield  {author} {\bibinfo {author} {\bibfnamefont {K.}~\bibnamefont
  {Schmidt-Hoberg}}, \bibinfo {author} {\bibfnamefont {F.}~\bibnamefont
  {Staub}}, \ and\ \bibinfo {author} {\bibfnamefont {M.~W.}\ \bibnamefont
  {Winkler}},\ }\href {\doibase 10.1016/j.physletb.2013.11.015} {\bibfield
  {journal} {\bibinfo  {journal} {Phys. Lett.}\ }\textbf {\bibinfo {volume}
  {B727}},\ \bibinfo {pages} {506} (\bibinfo {year} {2013})},\ \Eprint
  {http://arxiv.org/abs/1310.6752} {arXiv:1310.6752 [hep-ph]} \BibitemShut
  {NoStop}%
\bibitem [{\citenamefont {Piazza}\ and\ \citenamefont
  {Pospelov}(2010)}]{Piazza:2010ye}%
  \BibitemOpen
  \bibfield  {author} {\bibinfo {author} {\bibfnamefont {F.}~\bibnamefont
  {Piazza}}\ and\ \bibinfo {author} {\bibfnamefont {M.}~\bibnamefont
  {Pospelov}},\ }\href {\doibase 10.1103/PhysRevD.82.043533} {\bibfield
  {journal} {\bibinfo  {journal} {Phys. Rev.}\ }\textbf {\bibinfo {volume}
  {D82}},\ \bibinfo {pages} {043533} (\bibinfo {year} {2010})},\ \Eprint
  {http://arxiv.org/abs/1003.2313} {arXiv:1003.2313 [hep-ph]} \BibitemShut
  {NoStop}%
\bibitem [{\citenamefont {Kouvaris}\ \emph {et~al.}(2015)\citenamefont
  {Kouvaris}, \citenamefont {Shoemaker},\ and\ \citenamefont
  {Tuominen}}]{Kouvaris:2014uoa}%
  \BibitemOpen
  \bibfield  {author} {\bibinfo {author} {\bibfnamefont {C.}~\bibnamefont
  {Kouvaris}}, \bibinfo {author} {\bibfnamefont {I.~M.}\ \bibnamefont
  {Shoemaker}}, \ and\ \bibinfo {author} {\bibfnamefont {K.}~\bibnamefont
  {Tuominen}},\ }\href {\doibase 10.1103/PhysRevD.91.043519} {\bibfield
  {journal} {\bibinfo  {journal} {Phys. Rev.}\ }\textbf {\bibinfo {volume}
  {D91}},\ \bibinfo {pages} {043519} (\bibinfo {year} {2015})},\ \Eprint
  {http://arxiv.org/abs/1411.3730} {arXiv:1411.3730 [hep-ph]} \BibitemShut
  {NoStop}%
\bibitem [{\citenamefont {Kainulainen}\ \emph {et~al.}(2015)\citenamefont
  {Kainulainen}, \citenamefont {Tuominen},\ and\ \citenamefont
  {Vaskonen}}]{Kainulainen:2015sva}%
  \BibitemOpen
  \bibfield  {author} {\bibinfo {author} {\bibfnamefont {K.}~\bibnamefont
  {Kainulainen}}, \bibinfo {author} {\bibfnamefont {K.}~\bibnamefont
  {Tuominen}}, \ and\ \bibinfo {author} {\bibfnamefont {V.}~\bibnamefont
  {Vaskonen}},\ }\href@noop {} {\  (\bibinfo {year} {2015})},\ \Eprint
  {http://arxiv.org/abs/1507.04931} {arXiv:1507.04931 [hep-ph]} \BibitemShut
  {NoStop}%
\bibitem [{\citenamefont {Rajaraman}\ \emph {et~al.}(2015)\citenamefont
  {Rajaraman}, \citenamefont {Smolinsky},\ and\ \citenamefont
  {Tanedo}}]{Rajaraman:2015xka}%
  \BibitemOpen
  \bibfield  {author} {\bibinfo {author} {\bibfnamefont {A.}~\bibnamefont
  {Rajaraman}}, \bibinfo {author} {\bibfnamefont {J.}~\bibnamefont
  {Smolinsky}}, \ and\ \bibinfo {author} {\bibfnamefont {P.}~\bibnamefont
  {Tanedo}},\ }\href@noop {} {\  (\bibinfo {year} {2015})},\ \Eprint
  {http://arxiv.org/abs/1503.05919} {arXiv:1503.05919 [hep-ph]} \BibitemShut
  {NoStop}%
\bibitem [{\citenamefont {Hochberg}\ \emph {et~al.}(2014)\citenamefont
  {Hochberg}, \citenamefont {Kuflik}, \citenamefont {Volansky},\ and\
  \citenamefont {Wacker}}]{Hochberg:2014dra}%
  \BibitemOpen
  \bibfield  {author} {\bibinfo {author} {\bibfnamefont {Y.}~\bibnamefont
  {Hochberg}}, \bibinfo {author} {\bibfnamefont {E.}~\bibnamefont {Kuflik}},
  \bibinfo {author} {\bibfnamefont {T.}~\bibnamefont {Volansky}}, \ and\
  \bibinfo {author} {\bibfnamefont {J.~G.}\ \bibnamefont {Wacker}},\ }\href
  {\doibase 10.1103/PhysRevLett.113.171301} {\bibfield  {journal} {\bibinfo
  {journal} {Phys. Rev. Lett.}\ }\textbf {\bibinfo {volume} {113}},\ \bibinfo
  {pages} {171301} (\bibinfo {year} {2014})},\ \Eprint
  {http://arxiv.org/abs/1402.5143} {arXiv:1402.5143 [hep-ph]} \BibitemShut
  {NoStop}%
\bibitem [{\citenamefont {Clarke}\ \emph {et~al.}(2014)\citenamefont {Clarke},
  \citenamefont {Foot},\ and\ \citenamefont {Volkas}}]{Clarke:2013aya}%
  \BibitemOpen
  \bibfield  {author} {\bibinfo {author} {\bibfnamefont {J.~D.}\ \bibnamefont
  {Clarke}}, \bibinfo {author} {\bibfnamefont {R.}~\bibnamefont {Foot}}, \ and\
  \bibinfo {author} {\bibfnamefont {R.~R.}\ \bibnamefont {Volkas}},\ }\href
  {\doibase 10.1007/JHEP02(2014)123} {\bibfield  {journal} {\bibinfo  {journal}
  {JHEP}\ }\textbf {\bibinfo {volume} {02}},\ \bibinfo {pages} {123} (\bibinfo
  {year} {2014})},\ \Eprint {http://arxiv.org/abs/1310.8042} {arXiv:1310.8042
  [hep-ph]} \BibitemShut {NoStop}%
\bibitem [{\citenamefont {Gunion}\ \emph {et~al.}(2000)\citenamefont {Gunion},
  \citenamefont {Haber}, \citenamefont {Kane},\ and\ \citenamefont
  {Dawson}}]{Gunion:1989we}%
  \BibitemOpen
  \bibfield  {author} {\bibinfo {author} {\bibfnamefont {J.~F.}\ \bibnamefont
  {Gunion}}, \bibinfo {author} {\bibfnamefont {H.~E.}\ \bibnamefont {Haber}},
  \bibinfo {author} {\bibfnamefont {G.~L.}\ \bibnamefont {Kane}}, \ and\
  \bibinfo {author} {\bibfnamefont {S.}~\bibnamefont {Dawson}},\ }\href@noop {}
  {\bibfield  {journal} {\bibinfo  {journal} {Front. Phys.}\ }\textbf {\bibinfo
  {volume} {80}},\ \bibinfo {pages} {1} (\bibinfo {year} {2000})}\BibitemShut
  {NoStop}%
\bibitem [{\citenamefont {Voloshin}(1986)}]{Voloshin:1985tc}%
  \BibitemOpen
  \bibfield  {author} {\bibinfo {author} {\bibfnamefont {M.~B.}\ \bibnamefont
  {Voloshin}},\ }\href@noop {} {\bibfield  {journal} {\bibinfo  {journal} {Sov.
  J. Nucl. Phys.}\ }\textbf {\bibinfo {volume} {44}},\ \bibinfo {pages} {478}
  (\bibinfo {year} {1986})},\ \bibinfo {note} {[Yad.
  Fiz.44,738(1986)]}\BibitemShut {NoStop}%
\bibitem [{\citenamefont {Grinstein}\ \emph {et~al.}(1988)\citenamefont
  {Grinstein}, \citenamefont {Hall},\ and\ \citenamefont
  {Randall}}]{Grinstein:1988yu}%
  \BibitemOpen
  \bibfield  {author} {\bibinfo {author} {\bibfnamefont {B.}~\bibnamefont
  {Grinstein}}, \bibinfo {author} {\bibfnamefont {L.~J.}\ \bibnamefont {Hall}},
  \ and\ \bibinfo {author} {\bibfnamefont {L.}~\bibnamefont {Randall}},\ }\href
  {\doibase 10.1016/0370-2693(88)90916-1} {\bibfield  {journal} {\bibinfo
  {journal} {Phys. Lett.}\ }\textbf {\bibinfo {volume} {B211}},\ \bibinfo
  {pages} {363} (\bibinfo {year} {1988})}\BibitemShut {NoStop}%
\bibitem [{\citenamefont {Raby}\ and\ \citenamefont
  {West}(1988)}]{Raby:1988qf}%
  \BibitemOpen
  \bibfield  {author} {\bibinfo {author} {\bibfnamefont {S.}~\bibnamefont
  {Raby}}\ and\ \bibinfo {author} {\bibfnamefont {G.~B.}\ \bibnamefont
  {West}},\ }\href {\doibase 10.1103/PhysRevD.38.3488} {\bibfield  {journal}
  {\bibinfo  {journal} {Phys. Rev.}\ }\textbf {\bibinfo {volume} {D38}},\
  \bibinfo {pages} {3488} (\bibinfo {year} {1988})}\BibitemShut {NoStop}%
\bibitem [{\citenamefont {Donoghue}\ \emph {et~al.}(1990)\citenamefont
  {Donoghue}, \citenamefont {Gasser},\ and\ \citenamefont
  {Leutwyler}}]{Donoghue:1990xh}%
  \BibitemOpen
  \bibfield  {author} {\bibinfo {author} {\bibfnamefont {J.~F.}\ \bibnamefont
  {Donoghue}}, \bibinfo {author} {\bibfnamefont {J.}~\bibnamefont {Gasser}}, \
  and\ \bibinfo {author} {\bibfnamefont {H.}~\bibnamefont {Leutwyler}},\ }\href
  {\doibase 10.1016/0550-3213(90)90474-R} {\bibfield  {journal} {\bibinfo
  {journal} {Nucl. Phys.}\ }\textbf {\bibinfo {volume} {B343}},\ \bibinfo
  {pages} {341} (\bibinfo {year} {1990})}\BibitemShut {NoStop}%
\bibitem [{\citenamefont {Truong}\ and\ \citenamefont
  {Willey}(1989)}]{Truong:1989my}%
  \BibitemOpen
  \bibfield  {author} {\bibinfo {author} {\bibfnamefont {T.~N.}\ \bibnamefont
  {Truong}}\ and\ \bibinfo {author} {\bibfnamefont {R.~S.}\ \bibnamefont
  {Willey}},\ }\href {\doibase 10.1103/PhysRevD.40.3635} {\bibfield  {journal}
  {\bibinfo  {journal} {Phys. Rev.}\ }\textbf {\bibinfo {volume} {D40}},\
  \bibinfo {pages} {3635} (\bibinfo {year} {1989})}\BibitemShut {NoStop}%
\bibitem [{\citenamefont {Ade}\ \emph {et~al.}(2015)\citenamefont {Ade} \emph
  {et~al.}}]{Ade:2015xua}%
  \BibitemOpen
  \bibfield  {author} {\bibinfo {author} {\bibfnamefont {P.~A.~R.}\
  \bibnamefont {Ade}} \emph {et~al.} (\bibinfo {collaboration} {Planck}),\
  }\href@noop {} {\  (\bibinfo {year} {2015})},\ \Eprint
  {http://arxiv.org/abs/1502.01589} {arXiv:1502.01589 [astro-ph.CO]}
  \BibitemShut {NoStop}%
\bibitem [{\citenamefont {Nollett}\ and\ \citenamefont
  {Steigman}(2014)}]{Nollett:2013pwa}%
  \BibitemOpen
  \bibfield  {author} {\bibinfo {author} {\bibfnamefont {K.~M.}\ \bibnamefont
  {Nollett}}\ and\ \bibinfo {author} {\bibfnamefont {G.}~\bibnamefont
  {Steigman}},\ }\href {\doibase 10.1103/PhysRevD.89.083508} {\bibfield
  {journal} {\bibinfo  {journal} {Phys. Rev.}\ }\textbf {\bibinfo {volume}
  {D89}},\ \bibinfo {pages} {083508} (\bibinfo {year} {2014})},\ \Eprint
  {http://arxiv.org/abs/1312.5725} {arXiv:1312.5725 [astro-ph.CO]} \BibitemShut
  {NoStop}%
\bibitem [{\citenamefont {Aad}\ \emph {et~al.}(2016)\citenamefont {Aad} \emph
  {et~al.}}]{Aad:2015txa}%
  \BibitemOpen
  \bibfield  {author} {\bibinfo {author} {\bibfnamefont {G.}~\bibnamefont
  {Aad}} \emph {et~al.} (\bibinfo {collaboration} {ATLAS}),\ }\href {\doibase
  10.1007/JHEP01(2016)172} {\bibfield  {journal} {\bibinfo  {journal} {JHEP}\
  }\textbf {\bibinfo {volume} {01}},\ \bibinfo {pages} {172} (\bibinfo {year}
  {2016})},\ \Eprint {http://arxiv.org/abs/1508.07869} {arXiv:1508.07869
  [hep-ex]} \BibitemShut {NoStop}%
\bibitem [{\citenamefont {Curtin}\ \emph {et~al.}(2014)\citenamefont {Curtin}
  \emph {et~al.}}]{Curtin:2013fra}%
  \BibitemOpen
  \bibfield  {author} {\bibinfo {author} {\bibfnamefont {D.}~\bibnamefont
  {Curtin}} \emph {et~al.},\ }\href {\doibase 10.1103/PhysRevD.90.075004}
  {\bibfield  {journal} {\bibinfo  {journal} {Phys. Rev.}\ }\textbf {\bibinfo
  {volume} {D90}},\ \bibinfo {pages} {075004} (\bibinfo {year} {2014})},\
  \Eprint {http://arxiv.org/abs/1312.4992} {arXiv:1312.4992 [hep-ph]}
  \BibitemShut {NoStop}%
\bibitem [{\citenamefont {Dolan}\ \emph {et~al.}(2014)\citenamefont {Dolan},
  \citenamefont {McCabe}, \citenamefont {Kahlhoefer},\ and\ \citenamefont
  {Schmidt-Hoberg}}]{Dolan:2014ska}%
  \BibitemOpen
  \bibfield  {author} {\bibinfo {author} {\bibfnamefont {M.~J.}\ \bibnamefont
  {Dolan}}, \bibinfo {author} {\bibfnamefont {C.}~\bibnamefont {McCabe}},
  \bibinfo {author} {\bibfnamefont {F.}~\bibnamefont {Kahlhoefer}}, \ and\
  \bibinfo {author} {\bibfnamefont {K.}~\bibnamefont {Schmidt-Hoberg}},\
  }\href@noop {} {\  (\bibinfo {year} {2014})},\ \Eprint
  {http://arxiv.org/abs/1412.5174} {arXiv:1412.5174 [hep-ph]} \BibitemShut
  {NoStop}%
\bibitem [{\citenamefont {Ball}\ and\ \citenamefont
  {Zwicky}(2005)}]{Ball:2004ye}%
  \BibitemOpen
  \bibfield  {author} {\bibinfo {author} {\bibfnamefont {P.}~\bibnamefont
  {Ball}}\ and\ \bibinfo {author} {\bibfnamefont {R.}~\bibnamefont {Zwicky}},\
  }\href {\doibase 10.1103/PhysRevD.71.014015} {\bibfield  {journal} {\bibinfo
  {journal} {Phys. Rev.}\ }\textbf {\bibinfo {volume} {D71}},\ \bibinfo {pages}
  {014015} (\bibinfo {year} {2005})},\ \Eprint
  {http://arxiv.org/abs/hep-ph/0406232} {arXiv:hep-ph/0406232 [hep-ph]}
  \BibitemShut {NoStop}%
\bibitem [{\citenamefont {Olive}\ \emph {et~al.}(2014)\citenamefont {Olive}
  \emph {et~al.}}]{Agashe:2014kda}%
  \BibitemOpen
  \bibfield  {author} {\bibinfo {author} {\bibfnamefont {K.~A.}\ \bibnamefont
  {Olive}} \emph {et~al.} (\bibinfo {collaboration} {Particle Data Group}),\
  }\href {\doibase 10.1088/1674-1137/38/9/090001} {\bibfield  {journal}
  {\bibinfo  {journal} {Chin. Phys.}\ }\textbf {\bibinfo {volume} {C38}},\
  \bibinfo {pages} {090001} (\bibinfo {year} {2014})}\BibitemShut {NoStop}%
\bibitem [{\citenamefont {Lees}\ \emph {et~al.}(2013)\citenamefont {Lees} \emph
  {et~al.}}]{Lees:2013kla}%
  \BibitemOpen
  \bibfield  {author} {\bibinfo {author} {\bibfnamefont {J.~P.}\ \bibnamefont
  {Lees}} \emph {et~al.} (\bibinfo {collaboration} {BaBar}),\ }\href {\doibase
  10.1103/PhysRevD.87.112005} {\bibfield  {journal} {\bibinfo  {journal} {Phys.
  Rev.}\ }\textbf {\bibinfo {volume} {D87}},\ \bibinfo {pages} {112005}
  (\bibinfo {year} {2013})},\ \Eprint {http://arxiv.org/abs/1303.7465}
  {arXiv:1303.7465 [hep-ex]} \BibitemShut {NoStop}%
\bibitem [{\citenamefont {Marciano}\ and\ \citenamefont
  {Parsa}(1996)}]{Marciano:1996wy}%
  \BibitemOpen
  \bibfield  {author} {\bibinfo {author} {\bibfnamefont {W.}~\bibnamefont
  {Marciano}}\ and\ \bibinfo {author} {\bibfnamefont {Z.}~\bibnamefont
  {Parsa}},\ }\href {\doibase 10.1103/PhysRevD.53.R1} {\bibfield  {journal}
  {\bibinfo  {journal} {Phys.Rev.}\ }\textbf {\bibinfo {volume} {D53}},\
  \bibinfo {pages} {1} (\bibinfo {year} {1996})}\BibitemShut {NoStop}%
\bibitem [{\citenamefont {Artamonov}\ \emph {et~al.}(2008)\citenamefont
  {Artamonov} \emph {et~al.}}]{Artamonov:2008qb}%
  \BibitemOpen
  \bibfield  {author} {\bibinfo {author} {\bibfnamefont {A.~V.}\ \bibnamefont
  {Artamonov}} \emph {et~al.} (\bibinfo {collaboration} {E949}),\ }\href
  {\doibase 10.1103/PhysRevLett.101.191802} {\bibfield  {journal} {\bibinfo
  {journal} {Phys. Rev. Lett.}\ }\textbf {\bibinfo {volume} {101}},\ \bibinfo
  {pages} {191802} (\bibinfo {year} {2008})},\ \Eprint
  {http://arxiv.org/abs/0808.2459} {arXiv:0808.2459 [hep-ex]} \BibitemShut
  {NoStop}%
\bibitem [{\citenamefont {Ruggiero}(2013)}]{Ruggiero:2013oxa}%
  \BibitemOpen
  \bibfield  {author} {\bibinfo {author} {\bibfnamefont {G.}~\bibnamefont
  {Ruggiero}} (\bibinfo {collaboration} {NA62}),\ }\bibfield  {booktitle}
  {\emph {\bibinfo {booktitle} {{Proceedings, Kaon Physics International
  Conference (KAON13)}}},\ }\href@noop {} {\bibfield  {journal} {\bibinfo
  {journal} {PoS}\ }\textbf {\bibinfo {volume} {KAON13}},\ \bibinfo {pages}
  {032} (\bibinfo {year} {2013})}\BibitemShut {NoStop}%
\bibitem [{\citenamefont {Essig}\ \emph
  {et~al.}(2012{\natexlab{a}})\citenamefont {Essig}, \citenamefont
  {Manalaysay}, \citenamefont {Mardon}, \citenamefont {Sorensen},\ and\
  \citenamefont {Volansky}}]{Essig:2012yx}%
  \BibitemOpen
  \bibfield  {author} {\bibinfo {author} {\bibfnamefont {R.}~\bibnamefont
  {Essig}}, \bibinfo {author} {\bibfnamefont {A.}~\bibnamefont {Manalaysay}},
  \bibinfo {author} {\bibfnamefont {J.}~\bibnamefont {Mardon}}, \bibinfo
  {author} {\bibfnamefont {P.}~\bibnamefont {Sorensen}}, \ and\ \bibinfo
  {author} {\bibfnamefont {T.}~\bibnamefont {Volansky}},\ }\href {\doibase
  10.1103/PhysRevLett.109.021301} {\bibfield  {journal} {\bibinfo  {journal}
  {Phys.Rev.Lett.}\ }\textbf {\bibinfo {volume} {109}},\ \bibinfo {pages}
  {021301} (\bibinfo {year} {2012}{\natexlab{a}})},\ \Eprint
  {http://arxiv.org/abs/1206.2644} {arXiv:1206.2644 [astro-ph.CO]} \BibitemShut
  {NoStop}%
\bibitem [{\citenamefont {Essig}\ \emph
  {et~al.}(2012{\natexlab{b}})\citenamefont {Essig}, \citenamefont {Mardon},\
  and\ \citenamefont {Volansky}}]{Essig:2011nj}%
  \BibitemOpen
  \bibfield  {author} {\bibinfo {author} {\bibfnamefont {R.}~\bibnamefont
  {Essig}}, \bibinfo {author} {\bibfnamefont {J.}~\bibnamefont {Mardon}}, \
  and\ \bibinfo {author} {\bibfnamefont {T.}~\bibnamefont {Volansky}},\ }\href
  {\doibase 10.1103/PhysRevD.85.076007} {\bibfield  {journal} {\bibinfo
  {journal} {Phys.Rev.}\ }\textbf {\bibinfo {volume} {D85}},\ \bibinfo {pages}
  {076007} (\bibinfo {year} {2012}{\natexlab{b}})},\ \Eprint
  {http://arxiv.org/abs/1108.5383} {arXiv:1108.5383 [hep-ph]} \BibitemShut
  {NoStop}%
\bibitem [{\citenamefont {Shifman}\ \emph {et~al.}(1978)\citenamefont
  {Shifman}, \citenamefont {Vainshtein},\ and\ \citenamefont
  {Zakharov}}]{Shifman:1978zn}%
  \BibitemOpen
  \bibfield  {author} {\bibinfo {author} {\bibfnamefont {M.~A.}\ \bibnamefont
  {Shifman}}, \bibinfo {author} {\bibfnamefont {A.~I.}\ \bibnamefont
  {Vainshtein}}, \ and\ \bibinfo {author} {\bibfnamefont {V.~I.}\ \bibnamefont
  {Zakharov}},\ }\href {\doibase 10.1016/0370-2693(78)90481-1} {\bibfield
  {journal} {\bibinfo  {journal} {Phys. Lett.}\ }\textbf {\bibinfo {volume}
  {B78}},\ \bibinfo {pages} {443} (\bibinfo {year} {1978})}\BibitemShut
  {NoStop}%
\bibitem [{\citenamefont {Cirelli}\ \emph {et~al.}(2013)\citenamefont
  {Cirelli}, \citenamefont {Del~Nobile},\ and\ \citenamefont
  {Panci}}]{DelNobile:2013sia}%
  \BibitemOpen
  \bibfield  {author} {\bibinfo {author} {\bibfnamefont {M.}~\bibnamefont
  {Cirelli}}, \bibinfo {author} {\bibfnamefont {E.}~\bibnamefont {Del~Nobile}},
  \ and\ \bibinfo {author} {\bibfnamefont {P.}~\bibnamefont {Panci}},\ }\href
  {\doibase 10.1088/1475-7516/2013/10/019} {\bibfield  {journal} {\bibinfo
  {journal} {JCAP}\ }\textbf {\bibinfo {volume} {1310}},\ \bibinfo {pages}
  {019} (\bibinfo {year} {2013})},\ \Eprint {http://arxiv.org/abs/1307.5955}
  {arXiv:1307.5955 [hep-ph]} \BibitemShut {NoStop}%
\bibitem [{\citenamefont {Akerib}\ \emph {et~al.}(2014)\citenamefont {Akerib}
  \emph {et~al.}}]{Akerib:2013tjd}%
  \BibitemOpen
  \bibfield  {author} {\bibinfo {author} {\bibfnamefont {D.~S.}\ \bibnamefont
  {Akerib}} \emph {et~al.} (\bibinfo {collaboration} {LUX}),\ }\href {\doibase
  10.1103/PhysRevLett.112.091303} {\bibfield  {journal} {\bibinfo  {journal}
  {Phys. Rev. Lett.}\ }\textbf {\bibinfo {volume} {112}},\ \bibinfo {pages}
  {091303} (\bibinfo {year} {2014})},\ \Eprint {http://arxiv.org/abs/1310.8214}
  {arXiv:1310.8214 [astro-ph.CO]} \BibitemShut {NoStop}%
\bibitem [{\citenamefont {Agnese}\ \emph {et~al.}(2014)\citenamefont {Agnese}
  \emph {et~al.}}]{Agnese:2014aze}%
  \BibitemOpen
  \bibfield  {author} {\bibinfo {author} {\bibfnamefont {R.}~\bibnamefont
  {Agnese}} \emph {et~al.} (\bibinfo {collaboration} {SuperCDMS}),\ }\href
  {\doibase 10.1103/PhysRevLett.112.241302} {\bibfield  {journal} {\bibinfo
  {journal} {Phys. Rev. Lett.}\ }\textbf {\bibinfo {volume} {112}},\ \bibinfo
  {pages} {241302} (\bibinfo {year} {2014})},\ \Eprint
  {http://arxiv.org/abs/1402.7137} {arXiv:1402.7137 [hep-ex]} \BibitemShut
  {NoStop}%
\bibitem [{\citenamefont {Angloher}\ \emph {et~al.}(2015)\citenamefont
  {Angloher} \emph {et~al.}}]{Angloher:2015ewa}%
  \BibitemOpen
  \bibfield  {author} {\bibinfo {author} {\bibfnamefont {G.}~\bibnamefont
  {Angloher}} \emph {et~al.} (\bibinfo {collaboration} {CRESST}),\ }\href@noop
  {} {\  (\bibinfo {year} {2015})},\ \Eprint {http://arxiv.org/abs/1509.01515}
  {arXiv:1509.01515 [astro-ph.CO]} \BibitemShut {NoStop}%
\bibitem [{\citenamefont {Gerbier}\ \emph {et~al.}(2014)\citenamefont
  {Gerbier}, \citenamefont {Giomataris}, \citenamefont {Magnier}, \citenamefont
  {Dastgheibi}, \citenamefont {Gros} \emph {et~al.}}]{Gerbier:2014jwa}%
  \BibitemOpen
  \bibfield  {author} {\bibinfo {author} {\bibfnamefont {G.}~\bibnamefont
  {Gerbier}}, \bibinfo {author} {\bibfnamefont {I.}~\bibnamefont {Giomataris}},
  \bibinfo {author} {\bibfnamefont {P.}~\bibnamefont {Magnier}}, \bibinfo
  {author} {\bibfnamefont {A.}~\bibnamefont {Dastgheibi}}, \bibinfo {author}
  {\bibfnamefont {M.}~\bibnamefont {Gros}},  \emph {et~al.},\ }\href@noop {} {\
   (\bibinfo {year} {2014})},\ \Eprint {http://arxiv.org/abs/1401.7902}
  {arXiv:1401.7902 [astro-ph.IM]} \BibitemShut {NoStop}%
\bibitem [{\citenamefont {Cushman}\ \emph {et~al.}(2013)\citenamefont
  {Cushman}, \citenamefont {Galbiati}, \citenamefont {McKinsey}, \citenamefont
  {Robertson}, \citenamefont {Tait} \emph {et~al.}}]{Cushman:2013zza}%
  \BibitemOpen
  \bibfield  {author} {\bibinfo {author} {\bibfnamefont {P.}~\bibnamefont
  {Cushman}}, \bibinfo {author} {\bibfnamefont {C.}~\bibnamefont {Galbiati}},
  \bibinfo {author} {\bibfnamefont {D.}~\bibnamefont {McKinsey}}, \bibinfo
  {author} {\bibfnamefont {H.}~\bibnamefont {Robertson}}, \bibinfo {author}
  {\bibfnamefont {T.}~\bibnamefont {Tait}},  \emph {et~al.},\ }\href@noop {} {\
   (\bibinfo {year} {2013})},\ \Eprint {http://arxiv.org/abs/1310.8327}
  {arXiv:1310.8327 [hep-ex]} \BibitemShut {NoStop}%
\bibitem [{\citenamefont {Bergsma}\ \emph {et~al.}(1985)\citenamefont {Bergsma}
  \emph {et~al.}}]{Bergsma:1985qz}%
  \BibitemOpen
  \bibfield  {author} {\bibinfo {author} {\bibfnamefont {F.}~\bibnamefont
  {Bergsma}} \emph {et~al.} (\bibinfo {collaboration} {CHARM}),\ }\href
  {\doibase 10.1016/0370-2693(85)90400-9} {\bibfield  {journal} {\bibinfo
  {journal} {Phys. Lett.}\ }\textbf {\bibinfo {volume} {B157}},\ \bibinfo
  {pages} {458} (\bibinfo {year} {1985})}\BibitemShut {NoStop}%
\bibitem [{\citenamefont {Alekhin}\ \emph {et~al.}(2015)\citenamefont {Alekhin}
  \emph {et~al.}}]{Alekhin:2015byh}%
  \BibitemOpen
  \bibfield  {author} {\bibinfo {author} {\bibfnamefont {S.}~\bibnamefont
  {Alekhin}} \emph {et~al.},\ }\href@noop {} {\  (\bibinfo {year} {2015})},\
  \Eprint {http://arxiv.org/abs/1504.04855} {arXiv:1504.04855 [hep-ph]}
  \BibitemShut {NoStop}%
\bibitem [{\citenamefont {Aaij}\ \emph {et~al.}(2013)\citenamefont {Aaij} \emph
  {et~al.}}]{Aaij:2012vr}%
  \BibitemOpen
  \bibfield  {author} {\bibinfo {author} {\bibfnamefont {R.}~\bibnamefont
  {Aaij}} \emph {et~al.} (\bibinfo {collaboration} {LHCb}),\ }\href {\doibase
  10.1007/JHEP02(2013)105} {\bibfield  {journal} {\bibinfo  {journal} {JHEP}\
  }\textbf {\bibinfo {volume} {02}},\ \bibinfo {pages} {105} (\bibinfo {year}
  {2013})},\ \Eprint {http://arxiv.org/abs/1209.4284} {arXiv:1209.4284
  [hep-ex]} \BibitemShut {NoStop}%
\bibitem [{\citenamefont {Wei}\ \emph {et~al.}(2009)\citenamefont {Wei} \emph
  {et~al.}}]{Wei:2009zv}%
  \BibitemOpen
  \bibfield  {author} {\bibinfo {author} {\bibfnamefont {J.~T.}\ \bibnamefont
  {Wei}} \emph {et~al.} (\bibinfo {collaboration} {Belle}),\ }\href {\doibase
  10.1103/PhysRevLett.103.171801} {\bibfield  {journal} {\bibinfo  {journal}
  {Phys. Rev. Lett.}\ }\textbf {\bibinfo {volume} {103}},\ \bibinfo {pages}
  {171801} (\bibinfo {year} {2009})},\ \Eprint {http://arxiv.org/abs/0904.0770}
  {arXiv:0904.0770 [hep-ex]} \BibitemShut {NoStop}%
\bibitem [{\citenamefont {Alavi-Harati}\ \emph {et~al.}(2000)\citenamefont
  {Alavi-Harati} \emph {et~al.}}]{AlaviHarati:2000hs}%
  \BibitemOpen
  \bibfield  {author} {\bibinfo {author} {\bibfnamefont {A.}~\bibnamefont
  {Alavi-Harati}} \emph {et~al.} (\bibinfo {collaboration} {KTEV}),\ }\href
  {\doibase 10.1103/PhysRevLett.84.5279} {\bibfield  {journal} {\bibinfo
  {journal} {Phys. Rev. Lett.}\ }\textbf {\bibinfo {volume} {84}},\ \bibinfo
  {pages} {5279} (\bibinfo {year} {2000})},\ \Eprint
  {http://arxiv.org/abs/hep-ex/0001006} {arXiv:hep-ex/0001006 [hep-ex]}
  \BibitemShut {NoStop}%
\bibitem [{\citenamefont {Alavi-Harati}\ \emph {et~al.}(2004)\citenamefont
  {Alavi-Harati} \emph {et~al.}}]{AlaviHarati:2003mr}%
  \BibitemOpen
  \bibfield  {author} {\bibinfo {author} {\bibfnamefont {A.}~\bibnamefont
  {Alavi-Harati}} \emph {et~al.} (\bibinfo {collaboration} {KTeV}),\ }\href
  {\doibase 10.1103/PhysRevLett.93.021805} {\bibfield  {journal} {\bibinfo
  {journal} {Phys. Rev. Lett.}\ }\textbf {\bibinfo {volume} {93}},\ \bibinfo
  {pages} {021805} (\bibinfo {year} {2004})},\ \Eprint
  {http://arxiv.org/abs/hep-ex/0309072} {arXiv:hep-ex/0309072 [hep-ex]}
  \BibitemShut {NoStop}%
\bibitem [{\citenamefont {Djouadi}(2008)}]{Djouadi:2005gi}%
  \BibitemOpen
  \bibfield  {author} {\bibinfo {author} {\bibfnamefont {A.}~\bibnamefont
  {Djouadi}},\ }\href {\doibase 10.1016/j.physrep.2007.10.004} {\bibfield
  {journal} {\bibinfo  {journal} {Phys. Rept.}\ }\textbf {\bibinfo {volume}
  {457}},\ \bibinfo {pages} {1} (\bibinfo {year} {2008})},\ \Eprint
  {http://arxiv.org/abs/hep-ph/0503172} {arXiv:hep-ph/0503172 [hep-ph]}
  \BibitemShut {NoStop}%
\bibitem [{\citenamefont {Turner}(1988)}]{Turner:1987by}%
  \BibitemOpen
  \bibfield  {author} {\bibinfo {author} {\bibfnamefont {M.~S.}\ \bibnamefont
  {Turner}},\ }\href {\doibase 10.1103/PhysRevLett.60.1797} {\bibfield
  {journal} {\bibinfo  {journal} {Phys. Rev. Lett.}\ }\textbf {\bibinfo
  {volume} {60}},\ \bibinfo {pages} {1797} (\bibinfo {year}
  {1988})}\BibitemShut {NoStop}%
\bibitem [{\citenamefont {Frieman}\ \emph {et~al.}(1987)\citenamefont
  {Frieman}, \citenamefont {Dimopoulos},\ and\ \citenamefont
  {Turner}}]{Frieman:1987ui}%
  \BibitemOpen
  \bibfield  {author} {\bibinfo {author} {\bibfnamefont {J.~A.}\ \bibnamefont
  {Frieman}}, \bibinfo {author} {\bibfnamefont {S.}~\bibnamefont {Dimopoulos}},
  \ and\ \bibinfo {author} {\bibfnamefont {M.~S.}\ \bibnamefont {Turner}},\
  }\href {\doibase 10.1103/PhysRevD.36.2201} {\bibfield  {journal} {\bibinfo
  {journal} {Phys. Rev.}\ }\textbf {\bibinfo {volume} {D36}},\ \bibinfo {pages}
  {2201} (\bibinfo {year} {1987})}\BibitemShut {NoStop}%
\bibitem [{\citenamefont {Burrows}\ \emph {et~al.}(1989)\citenamefont
  {Burrows}, \citenamefont {Turner},\ and\ \citenamefont
  {Brinkmann}}]{Burrows:1988ah}%
  \BibitemOpen
  \bibfield  {author} {\bibinfo {author} {\bibfnamefont {A.}~\bibnamefont
  {Burrows}}, \bibinfo {author} {\bibfnamefont {M.~S.}\ \bibnamefont {Turner}},
  \ and\ \bibinfo {author} {\bibfnamefont {R.~P.}\ \bibnamefont {Brinkmann}},\
  }\href {\doibase 10.1103/PhysRevD.39.1020} {\bibfield  {journal} {\bibinfo
  {journal} {Phys. Rev.}\ }\textbf {\bibinfo {volume} {D39}},\ \bibinfo {pages}
  {1020} (\bibinfo {year} {1989})}\BibitemShut {NoStop}%
\bibitem [{\citenamefont {Essig}\ \emph {et~al.}(2010)\citenamefont {Essig},
  \citenamefont {Harnik}, \citenamefont {Kaplan},\ and\ \citenamefont
  {Toro}}]{Essig:2010gu}%
  \BibitemOpen
  \bibfield  {author} {\bibinfo {author} {\bibfnamefont {R.}~\bibnamefont
  {Essig}}, \bibinfo {author} {\bibfnamefont {R.}~\bibnamefont {Harnik}},
  \bibinfo {author} {\bibfnamefont {J.}~\bibnamefont {Kaplan}}, \ and\ \bibinfo
  {author} {\bibfnamefont {N.}~\bibnamefont {Toro}},\ }\href {\doibase
  10.1103/PhysRevD.82.113008} {\bibfield  {journal} {\bibinfo  {journal} {Phys.
  Rev.}\ }\textbf {\bibinfo {volume} {D82}},\ \bibinfo {pages} {113008}
  (\bibinfo {year} {2010})},\ \Eprint {http://arxiv.org/abs/1008.0636}
  {arXiv:1008.0636 [hep-ph]} \BibitemShut {NoStop}%
\bibitem [{\citenamefont {Ishizuka}\ and\ \citenamefont
  {Yoshimura}(1990)}]{Ishizuka:1989ts}%
  \BibitemOpen
  \bibfield  {author} {\bibinfo {author} {\bibfnamefont {N.}~\bibnamefont
  {Ishizuka}}\ and\ \bibinfo {author} {\bibfnamefont {M.}~\bibnamefont
  {Yoshimura}},\ }\href {\doibase 10.1143/PTP.84.233} {\bibfield  {journal}
  {\bibinfo  {journal} {Prog. Theor. Phys.}\ }\textbf {\bibinfo {volume}
  {84}},\ \bibinfo {pages} {233} (\bibinfo {year} {1990})}\BibitemShut
  {NoStop}%
\bibitem [{\citenamefont {Kaplan}(1992)}]{Kaplan:1991ah}%
  \BibitemOpen
  \bibfield  {author} {\bibinfo {author} {\bibfnamefont {D.~B.}\ \bibnamefont
  {Kaplan}},\ }\href {\doibase 10.1103/PhysRevLett.68.741} {\bibfield
  {journal} {\bibinfo  {journal} {Phys. Rev. Lett.}\ }\textbf {\bibinfo
  {volume} {68}},\ \bibinfo {pages} {741} (\bibinfo {year} {1992})}\BibitemShut
  {NoStop}%
\bibitem [{\citenamefont {Nussinov}(1985)}]{Nussinov:1985xr}%
  \BibitemOpen
  \bibfield  {author} {\bibinfo {author} {\bibfnamefont {S.}~\bibnamefont
  {Nussinov}},\ }\href {\doibase 10.1016/0370-2693(85)90689-6} {\bibfield
  {journal} {\bibinfo  {journal} {Phys. Lett.}\ }\textbf {\bibinfo {volume}
  {B165}},\ \bibinfo {pages} {55} (\bibinfo {year} {1985})}\BibitemShut
  {NoStop}%
\bibitem [{\citenamefont {Barr}(1991)}]{Barr:1991qn}%
  \BibitemOpen
  \bibfield  {author} {\bibinfo {author} {\bibfnamefont {S.~M.}\ \bibnamefont
  {Barr}},\ }\href {\doibase 10.1103/PhysRevD.44.3062} {\bibfield  {journal}
  {\bibinfo  {journal} {Phys. Rev.}\ }\textbf {\bibinfo {volume} {D44}},\
  \bibinfo {pages} {3062} (\bibinfo {year} {1991})}\BibitemShut {NoStop}%
\bibitem [{\citenamefont {Barr}\ \emph {et~al.}(1990)\citenamefont {Barr},
  \citenamefont {Chivukula},\ and\ \citenamefont {Farhi}}]{Barr:1990ca}%
  \BibitemOpen
  \bibfield  {author} {\bibinfo {author} {\bibfnamefont {S.~M.}\ \bibnamefont
  {Barr}}, \bibinfo {author} {\bibfnamefont {R.~S.}\ \bibnamefont {Chivukula}},
  \ and\ \bibinfo {author} {\bibfnamefont {E.}~\bibnamefont {Farhi}},\ }\href
  {\doibase 10.1016/0370-2693(90)91661-T} {\bibfield  {journal} {\bibinfo
  {journal} {Phys. Lett.}\ }\textbf {\bibinfo {volume} {B241}},\ \bibinfo
  {pages} {387} (\bibinfo {year} {1990})}\BibitemShut {NoStop}%
\bibitem [{\citenamefont {Kaplan}\ \emph {et~al.}(2009)\citenamefont {Kaplan},
  \citenamefont {Luty},\ and\ \citenamefont {Zurek}}]{Kaplan:2009ag}%
  \BibitemOpen
  \bibfield  {author} {\bibinfo {author} {\bibfnamefont {D.~E.}\ \bibnamefont
  {Kaplan}}, \bibinfo {author} {\bibfnamefont {M.~A.}\ \bibnamefont {Luty}}, \
  and\ \bibinfo {author} {\bibfnamefont {K.~M.}\ \bibnamefont {Zurek}},\ }\href
  {\doibase 10.1103/PhysRevD.79.115016} {\bibfield  {journal} {\bibinfo
  {journal} {Phys. Rev.}\ }\textbf {\bibinfo {volume} {D79}},\ \bibinfo {pages}
  {115016} (\bibinfo {year} {2009})},\ \Eprint {http://arxiv.org/abs/0901.4117}
  {arXiv:0901.4117 [hep-ph]} \BibitemShut {NoStop}%
\bibitem [{\citenamefont {Graesser}\ \emph {et~al.}(2011)\citenamefont
  {Graesser}, \citenamefont {Shoemaker},\ and\ \citenamefont
  {Vecchi}}]{Graesser:2011wi}%
  \BibitemOpen
  \bibfield  {author} {\bibinfo {author} {\bibfnamefont {M.~L.}\ \bibnamefont
  {Graesser}}, \bibinfo {author} {\bibfnamefont {I.~M.}\ \bibnamefont
  {Shoemaker}}, \ and\ \bibinfo {author} {\bibfnamefont {L.}~\bibnamefont
  {Vecchi}},\ }\href {\doibase 10.1007/JHEP10(2011)110} {\bibfield  {journal}
  {\bibinfo  {journal} {JHEP}\ }\textbf {\bibinfo {volume} {10}},\ \bibinfo
  {pages} {110} (\bibinfo {year} {2011})},\ \Eprint
  {http://arxiv.org/abs/1103.2771} {arXiv:1103.2771 [hep-ph]} \BibitemShut
  {NoStop}%
\bibitem [{\citenamefont {Lin}\ \emph {et~al.}(2012)\citenamefont {Lin},
  \citenamefont {Yu},\ and\ \citenamefont {Zurek}}]{Lin:2011gj}%
  \BibitemOpen
  \bibfield  {author} {\bibinfo {author} {\bibfnamefont {T.}~\bibnamefont
  {Lin}}, \bibinfo {author} {\bibfnamefont {H.-B.}\ \bibnamefont {Yu}}, \ and\
  \bibinfo {author} {\bibfnamefont {K.~M.}\ \bibnamefont {Zurek}},\ }\href
  {\doibase 10.1103/PhysRevD.85.063503} {\bibfield  {journal} {\bibinfo
  {journal} {Phys. Rev.}\ }\textbf {\bibinfo {volume} {D85}},\ \bibinfo {pages}
  {063503} (\bibinfo {year} {2012})},\ \Eprint {http://arxiv.org/abs/1111.0293}
  {arXiv:1111.0293 [hep-ph]} \BibitemShut {NoStop}%
\bibitem [{\citenamefont {Tucker-Smith}\ and\ \citenamefont
  {Weiner}(2001)}]{TuckerSmith:2001hy}%
  \BibitemOpen
  \bibfield  {author} {\bibinfo {author} {\bibfnamefont {D.}~\bibnamefont
  {Tucker-Smith}}\ and\ \bibinfo {author} {\bibfnamefont {N.}~\bibnamefont
  {Weiner}},\ }\href {\doibase 10.1103/PhysRevD.64.043502} {\bibfield
  {journal} {\bibinfo  {journal} {Phys. Rev.}\ }\textbf {\bibinfo {volume}
  {D64}},\ \bibinfo {pages} {043502} (\bibinfo {year} {2001})},\ \Eprint
  {http://arxiv.org/abs/hep-ph/0101138} {arXiv:hep-ph/0101138 [hep-ph]}
  \BibitemShut {NoStop}%
\bibitem [{\citenamefont {Bai}\ and\ \citenamefont {Tait}(2012)}]{Bai:2011jg}%
  \BibitemOpen
  \bibfield  {author} {\bibinfo {author} {\bibfnamefont {Y.}~\bibnamefont
  {Bai}}\ and\ \bibinfo {author} {\bibfnamefont {T.~M.~P.}\ \bibnamefont
  {Tait}},\ }\href {\doibase 10.1016/j.physletb.2012.03.011} {\bibfield
  {journal} {\bibinfo  {journal} {Phys. Lett.}\ }\textbf {\bibinfo {volume}
  {B710}},\ \bibinfo {pages} {335} (\bibinfo {year} {2012})},\ \Eprint
  {http://arxiv.org/abs/1109.4144} {arXiv:1109.4144 [hep-ph]} \BibitemShut
  {NoStop}%
\bibitem [{\citenamefont {Izaguirre}\ \emph
  {et~al.}(2015{\natexlab{c}})\citenamefont {Izaguirre}, \citenamefont
  {Krnjaic},\ and\ \citenamefont {Shuve}}]{Izaguirre:2015zva}%
  \BibitemOpen
  \bibfield  {author} {\bibinfo {author} {\bibfnamefont {E.}~\bibnamefont
  {Izaguirre}}, \bibinfo {author} {\bibfnamefont {G.}~\bibnamefont {Krnjaic}},
  \ and\ \bibinfo {author} {\bibfnamefont {B.}~\bibnamefont {Shuve}},\
  }\href@noop {} {\  (\bibinfo {year} {2015}{\natexlab{c}})},\ \Eprint
  {http://arxiv.org/abs/1508.03050} {arXiv:1508.03050 [hep-ph]} \BibitemShut
  {NoStop}%
\bibitem [{\citenamefont {Batell}\ \emph {et~al.}(2012)\citenamefont {Batell},
  \citenamefont {McKeen},\ and\ \citenamefont {Pospelov}}]{Batell:2012mj}%
  \BibitemOpen
  \bibfield  {author} {\bibinfo {author} {\bibfnamefont {B.}~\bibnamefont
  {Batell}}, \bibinfo {author} {\bibfnamefont {D.}~\bibnamefont {McKeen}}, \
  and\ \bibinfo {author} {\bibfnamefont {M.}~\bibnamefont {Pospelov}},\ }\href
  {\doibase 10.1007/JHEP10(2012)104} {\bibfield  {journal} {\bibinfo  {journal}
  {JHEP}\ }\textbf {\bibinfo {volume} {10}},\ \bibinfo {pages} {104} (\bibinfo
  {year} {2012})},\ \Eprint {http://arxiv.org/abs/1207.6252} {arXiv:1207.6252
  [hep-ph]} \BibitemShut {NoStop}%
\bibitem [{\citenamefont {Gondolo}\ and\ \citenamefont
  {Gelmini}(1991)}]{Gondolo:1990dk}%
  \BibitemOpen
  \bibfield  {author} {\bibinfo {author} {\bibfnamefont {P.}~\bibnamefont
  {Gondolo}}\ and\ \bibinfo {author} {\bibfnamefont {G.}~\bibnamefont
  {Gelmini}},\ }\href {\doibase 10.1016/0550-3213(91)90438-4} {\bibfield
  {journal} {\bibinfo  {journal} {Nucl. Phys.}\ }\textbf {\bibinfo {volume}
  {B360}},\ \bibinfo {pages} {145} (\bibinfo {year} {1991})}\BibitemShut
  {NoStop}%
\bibitem [{\citenamefont {Kolb}\ and\ \citenamefont
  {Turner}(1990)}]{Kolb:1990vq}%
  \BibitemOpen
  \bibfield  {author} {\bibinfo {author} {\bibfnamefont {E.~W.}\ \bibnamefont
  {Kolb}}\ and\ \bibinfo {author} {\bibfnamefont {M.~S.}\ \bibnamefont
  {Turner}},\ }\href@noop {} {\bibfield  {journal} {\bibinfo  {journal} {Front.
  Phys.}\ }\textbf {\bibinfo {volume} {69}},\ \bibinfo {pages} {1} (\bibinfo
  {year} {1990})}\BibitemShut {NoStop}%
\end{thebibliography}%

\bigskip


\section*{Appendix A: Higgs-Mediator Mixing} 
In this appendix, we review the mixing ansatz chosen in the body of the paper. Following the notation in \cite{Batell:2012mj}, above the scale of electroweak symmetry breaking, the most general renormalizable scalar potential for a SM Higgs doublet and scalar singlet is 
\be\hspace{-0.1cm}
&&V = V_{H}+ V_\Phi + V_{\Phi H}~,~~
\ee
where the contributions are 
\be  \label{eq:potential-couplings}
&&V_H =  \mu_H^2 H^\dagger H + \lambda_H (H^\dagger H)^2  \\ 
&&V_\Phi =  B_\Phi \Phi+  \frac{\mu_\Phi^2}{2} \Phi^2 + \frac{A_\Phi}{3} \Phi^3 + \frac{\lambda_\Phi}{4} \Phi^4\\
&&V_{\Phi H}  = (   A_{\tiny \Phi H} \Phi  +  \lambda_{\Phi H} \Phi^2 )      H^\dagger H ~~.~~
\ee
For simplicity and without loss of essential generality, we choose $B_{\Phi} = A_{\Phi H} v^2/2$ to ensure that $\Phi$ does not receive a VEV. After electroweak
symmetry breaking, $\Phi$ mixes with the scalar component of $H$ and the mass matrix is 
\begin{equation} 
{\bf M}^2
= 
\left( 
\begin{array}{cc}
2 \lambda_H v^2 & A_{\Phi H} v    \\
A_{\Phi H} v &  \mu_\Phi^2
\end{array}
\right)
, ~~~
\end{equation}
which is diagonalized by the rotation 
\begin{equation} 
\left( 
\begin{array}{c}
 H \\
 \Phi
\end{array} 
\right)
= 
\left( 
\begin{array}{cc}
\cos\theta &  \sin\theta \\
-\sin\theta &  \cos\theta
\end{array}
\right)
\left( 
\begin{array}{c}
 h \\
 \phi
\end{array} 
\right), ~~~
\end{equation}
where $h$ is the physical Higgs boson, $\phi$ is the dark sector mediator,
and 
\be
\tan 2\theta = \frac{2 A_{\Phi H} v }{\mu_\Phi^2 -2\lambda_{H} v^2}~. 
\ee
The mass eigenvalues in this setup are given by
\be
m^2_{h,\phi} = \frac{1}{2} \left(  \mu_\Phi^2 + 2 \lambda_{H} v^2   \pm  \sqrt{F}  \right)~,~~~
\ee
where we have defined 
\be
F \equiv   \mu_\Phi^4 + 4 A_{H \Phi}^2 v^2  + 4 v^4 \lambda_{H}^2  - 4 \lambda_{H}m_\Phi^2 v^2   ~~.~~~
\ee
Note that for appropriate choices of potential couplings in Eq.~(\ref{eq:potential-couplings}) it is always possible to engineer 
 a sufficiently light scalar $\phi$ in the mass eiegenbasis, though fine tuning may be required. 
 

\section*{Appendix B: Relic Density Computation}
To compute the relic density curve for the direct annihilation scenario, we begin by evaluating the total cross section $\sigma(s)$ for $\chi \chi \to f\bar f$ annihilation 
which recovers  Eq.~(\ref{eq:directannihilation})  in the nonrelativistic limit. 
Performing the thermal average \cite{Gondolo:1990dk}
\be
\langle \sigma |v|   \rangle_{ \chi \bar \chi  \to f\bar f   } \! =  \frac{1}{N_\chi} \int_{4m_\chi^2}^\infty  \! \! \! ds  \, \sigma(s) (s-4m_\chi^2)\sqrt{s} K_1 \! \left( \frac{\! \sqrt{s}}{T}\right) ,~~~~~~~~
\ee
where  $N_\chi = 8 m_\chi^4 T K_2^2 (m_\chi/T)$  and $K_n$ is a Bessel function of the $n^{\rm th.}$ kind. The annihilation cross section is 
\be
\sigma(s)_{\chi \chi \to f\bar f} = \frac{ g_\chi^2 g_f(s)^2 \, s}{16 \pi (s-m_\phi^2)^2}   \sqrt{  1-\frac{4m_f^2}{s}  } \left(1-\frac{4m_\chi^2}{s}\right)^{3/2}   \!\!\!\! ,~~~~~~~~~
\ee
where the mediator-SM coupling depends on the available hadronic thresholds accessible  for a given value of $s$. Near $\Lambda_{\rm QCD}$, we model this coupling 
as  \cite{Clarke:2013aya} 
\be
  g_{f}(s) \simeq   \sin\theta \sqrt{   \frac{8 \pi}{m_h} \Gamma(h \to {\rm SM})  }  ~ \biggr  |_{m_h = \sqrt{s}}~~~,
\ee
and have checked that for $\sqrt{s}$ away from $\Lambda_{\rm QCD}$, where the final states consist only of elementary particles, this procedure matches onto
the analytical result computed using only SM yukawa couplings. 

Following the procedure in \cite{Kolb:1990vq}, the relic abundance of $\chi$ is
\be
\Omega_\chi h^2 = 1.07 \times 10^9    \,      \frac{\sqrt{ g_{*}   }}{g_{*,s}}     \frac{ (n+1)  \, x_f \,  \GeV^{-1} }{    m_{Pl} \,  \langle\sigma |v|\rangle_f}~,~~
\ee
where $n = 0,1$ for $s$ and $p$ wave annihilation, $ g_{*}$ and  $g_{*,s}$ are respectively the relativistic and entropic degrees of freedom,  $x \equiv m_\chi/T$, and an $f$ subscript denotes a freeze out value. 

\bigskip


\section*{Appendix C: Thermalization Criteria}
To produce DM thermally in equilibrium, at minimum we require the production rate to exceed the expansion rate at some point in the early universe, $n_f(T) \langle \sigma |v| \rangle_{f\bar f \to \chi \bar \chi} \gsim H(T)$, where $n_f$ is the fermion number density 
and $H(T) \simeq 1.66 \sqrt{g_*} T^2/m_{Pl}$. Per SM fermion species  $f$, the cross section for annihilation is 
\be
\sigma(s)_{ f\bar f \to \chi \bar \chi} = \frac{ g_\chi^2 g_f^2 \, s}{16 \pi (s-m_\phi^2)^2}   \sqrt{  1-\frac{4m_\chi^2}{s}  } \left(1-\frac{4m_f^2}{s}\right)^{3/2}   \!\! \!\!,~~~~~~~~
\ee
The thermal average is 
\be
\langle \sigma |v|   \rangle_{ f\bar f \to \chi \bar \chi} \! =  \frac{1}{N_f} \int_{4m_\chi^2}^\infty  \! \! \! ds  \, \sigma(s) (s-4m_f^2)\sqrt{s} K_1 \! \left( \frac{\! \sqrt{s}}{T}\right) ,~~~~~~~~~~
\ee
where $N_f= 8 m_f^4 T K_2^2 (m_f/T)$. At very high temperatures, the number densities scale as $n_f(T) \langle\sigma |v|\rangle \propto T$, whereas $H(T)\propto T^2$ so as the universe cools, the thermal production rate increases relative to Hubble. In the opposite, non relativistic regime, the number density falls exponentially $n_f(T) \propto \exp(-m_f/T)$ so the rate decreases sharply relative to Hubble, which still scales as $T^2$ during radiation domination. The total rate for comparison with Hubble is 
\be
\Gamma_{{\rm SM} \to \chi \bar \chi} = \sum_f n_f(T) \langle \sigma |v| \rangle_{f\bar f\to \chi \bar \chi}~,~~
\ee
which we compare with the Hubble rate $\Gamma_{{\rm SM} \to \chi \bar \chi} $ near $T\sim m_t$ because this is the temperature at which the leading production cross section ($t\bar t \to \chi \bar \chi$) no longer scales as $T^{-2}$ but acquires $m_t$ dependence from the propagator as $s \approx 4 m_t^2$.


\section*{ Appendix D: Three Body Meson Decays} 
For $m_\phi > m_{B^+} - m_{K^+}$, it is still possible for DM to contribute to 
rare B decays through $B\to K \bar \chi \chi$ via virtual $\phi$ exchange. The width for this process can be computed
as a  convolution of 2 body processes, treating the intermediate $\phi$ propagator
as a Breit-Wigner function 
\be \label{eq:breit}
\hspace{-0.5cm}\Gamma_{B^+\to K^+\chi \bar \chi} &=& \int_{4m_\chi^2}^{ \, 0.3 \,m^2_{B} } dq^2 \Gamma_{B^+\to K^+ \phi}(q^2) F(q^2) ~,~ 
\ee 
where we define 
\be 
&&~~~~~~~ F(q^2) = \frac{   m_\phi \Gamma_{\phi}/\pi }{ (q^2 - m_\phi^2)^2 +m_\phi^2 \Gamma^2_{\phi}}~, \\ 
 &&\Gamma_{\phi}\equiv  \Gamma(\phi \to \chi \bar \chi) = \frac{    g_\chi^2 m_\phi }{8\pi} \left(1- \frac{  4m_\chi^2}{m_\phi^2} \right)^{3/2}~,~
 \ee
such that $F$ is normalized to recover a delta function in the vicinity of $m_\phi$ as $\Gamma_\phi \to 0$ limit. The upper integration limit in Eq.~(\ref{eq:breit}), $ q^2 =0.3 \, m_B^2,$
is chosen in accordance with the cut imposed in\cite{Lees:2013kla}. 
An analogous expression
is used to compute the width for $K^+\to \pi^+ \chi\bar \chi$ using the cuts in \cite{Artamonov:2008qb}, which constitutes the right most region of the orange shaded contour in Fig. 2.

\bigskip
\bigskip
\bigskip
\end{document}